\documentclass[lettersize,journal]{IEEEtran}
\usepackage{amsmath,amsfonts}
\usepackage{algorithmic}
\usepackage{algorithm}
\usepackage{array}
\usepackage[caption=false,font=normalsize,labelfont=sf,textfont=sf]{subfig}
\usepackage{textcomp}
\usepackage{stfloats}
\usepackage{url}
\usepackage{verbatim}
\usepackage{graphicx}
\usepackage{cite}
\usepackage{booktabs}
\usepackage{multirow}
\usepackage{subfig}
\usepackage{float}

\newcounter{mytempeqncnt}

\begin{document}

\title{Densifying MIMO: Channel Modeling, Physical Constraints, and Performance Evaluation for Holographic Communications}

\author{
	Yongxi Liu,~Ming Zhang,~\IEEEmembership{Member,~IEEE},~Tengjiao Wang,~\IEEEmembership{Member,~IEEE},
	
	~Anxue Zhang,~and Mérouane Debbah,~\IEEEmembership{Fellow,~IEEE}

	\thanks{This work was supported in part by the Shaanxi Key Laboratory of Deep Space Exploration Intelligent Information Technology under Grant No. 2021SYS-04. \textit{(Corresponding authors: Ming Zhang, Tengjiao Wang.)}}
	\thanks{Y. Liu, M. Zhang, and A. Zhang are with the School of Information and Communications Engineering, Xi'an Jiaotong University, Xi'an 710049, China (e-mail: liu\_xii@foxmail.com, ming20.zhang@xjtu.edu.cn, anxuezhang@xjtu.edu.cn).}
	\thanks{T. Wang is with the the Department of Wireless Network RAN Research, Huawei Technologies CO., Ltd, Shanghai 201206, China (e-mail: wangtengjiao6@huawei.com).}
	\thanks{M. Debbah is with KU 6G Research Center, Khalifa University of Science and Technology, P O Box 127788, Abu Dhabi, UAE (email: merouane.debbah@ku.ac.ae) and also with CentraleSupelec, University Paris-Saclay, 91192 Gif-sur-Yvette, France.}
}

\maketitle

\begin{abstract}
	
	As the backbone of the fifth-generation (5G) cellular network, massive multiple-input multiple-output (MIMO) encounters a significant challenge in practical applications: how to deploy a large number of antenna elements within limited spaces. Recently, holographic communication has emerged as a potential solution to this issue. It employs dense antenna arrays and provides a tractable model. Nevertheless, some challenges must be addressed to actualize this innovative concept. One is the mutual coupling among antenna elements within an array. When the element spacing is small, near-field coupling becomes the dominant factor that strongly restricts the array performance. Another is the polarization of electromagnetic waves. As an intrinsic property, it was not fully considered in the previous channel modeling of holographic communication. The third is the lack of real-world experiments to show the potential and possible defects of a holographic communication system. In this paper, we propose an electromagnetic channel model based on the characteristics of electromagnetic waves. This model encompasses the impact of mutual coupling in the transceiver sides and the depolarization in the propagation environment. Furthermore, by approximating an infinite array, the performance restrictions of large-scale dense antenna arrays are also studied theoretically to exploit the potential of the proposed channel. In addition, numerical simulations and a channel measurement experiment are conducted. The findings reveal that within limited spaces, the coupling effect, particularly for element spacing smaller than half of the wavelength, is the primary factor leading to the inflection point for the performance of holographic communications.
	
\end{abstract}


\IEEEpeerreviewmaketitle

\section{Introduction}

In the past few decades, with the rapid progress of mobile communication technology, people interact with more and more mobile terminals, leading to a larger demand for information transmission \cite{Chettri2020}. The use of 5G networks provides higher bandwidth and lower latency, and has attracted extensive attention. Under the international specification for 5G, engineers have been dedicated to developing the key technologies that improve network performance \cite{Boccardi2014,3GPP38901}. Among them, massive MIMO is considered to be one of the most promising technologies \cite{Marzetta2010,Larsson2014,Swindlehurst2014}. It utilizes spatial diversity, multiplexing, and beamforming, providing significant improvements in both spectral efficiency and network coverage \cite{Lu2014}. By deploying a large number of antenna elements in the base station (BS) side, massive MIMO effectively enhances network performance and facilitates the support of more user terminals (UTs), particularly in densely populated areas. As the antenna array gets larger and contains more elements, the communication channel exhibits deterministic characteristics, e.g., channel hardening and favorable propagation \cite{Bjornson2017}, which reduce the overhead of channel estimation and power allocation. However, the deployment of a larger number of elements in the antenna array leads to an inevitable increase in its size, making it challenging to meet the installation size of the base stations. Moreover, the number of antennas beyond a certain threshold causes increased hardware complexity and elevated power consumption, which is not allowable from an energy efficiency perspective \cite[Ch.~5]{Bjornson2017}. To address these practical limitations, researchers are devoted to exploring the next breakthrough beyond massive MIMO.

Focusing on the electromagnetic limitation behind massive MIMO communications, researchers are trying to combine the theoretical limits of Shannon’s information theory and Maxwell’s electromagnetic theory \cite{Loyka2004}. On the theoretical aspect, the wave theory of information helps reveal the intrinsic properties of mobile communications \cite{Ivrlac2014,Franceschetti2018}. Moreover, antenna miniaturization and the development of the intelligent reflecting surface (IRS) provide hardware solutions to deploy closed-space elements in an array \cite{Wu2019,Haraz2014}. In \cite{Pizzo2020Degrees,Pizzo2022Nyquist}, the authors analyze the degrees of freedom (DoF) of electromagnetic fields using the Nyquist sampling theorem in the wavenumber domain. The conclusions give rise to the study of a communication system between dense antenna arrays, named holographic MIMO communications \cite{Pizzo2020Spatially,Gong2023}. Given a limited space, if the element spacing is shortened, more elements can be deployed into the array, providing more sub-channels. Subsequently, in \cite{Pizzo2022Fourier,Pizzo2022Spatial}, the authors propose the holographic channel model under the non-line-of-sight (NLOS) scenario based on the Fourier plane-wave series expansion of electromagnetic waves. This model establishes the response relationship between the current density in the source array and the electric field in the receive array. Simulation shows that the DoF of holographic communication is fully determined by the size of transceiver arrays\cite{Pizzo2020Degrees}. And using equal power allocation in the wavenumber domain, channel capacity can increase boundlessly as the element spacing of transceiver arrays gets closer \cite{Pizzo2022Fourier}. This conclusion, however, is tenable by ignoring the effect of mutual coupling. In \cite{Sun2022,Yuan2023}, the effect of mutual coupling on the spatial DoF of holographic communication is investigated, but how it is related to system performance remains to be explored, and thus there is an urgent need for an accurate channel model that includes the coupling effect and other physical constraints. In \cite{Wei2022,Sha2022}, the authors extend the holographic channel model into the multi-user scenario and improve the scalability of holographic communications. Furthermore, in \cite{Akrout2023}, a circuit-based model is proposed to exploit the mutual coupling inside dense antenna arrays composed of Chu's minimum scattering antennas. Nevertheless, the results are confined to this special antenna form, and the considerations regarding the limitations of antenna radiation and matching networks are not sufficient. Hence, it needs to be investigated whether these idealized models fit into real-world communications.

In our previous studies \cite{Yongxi2022,Tengjiao2022}, we developed a modified channel model under scalar wave communication that accounts for mutual coupling among antenna elements. Nonetheless, the polarization of electromagnetic waves was not considered in this model. Furthermore, according to coupling and radiation theory, a finite-size dense antenna array exhibits certain constraints, including severe inter-element coupling and limited angular resolution. These factors, however, are of great importance in communication systems.

This paper aims to address these challenges within the context of holographic MIMO communications. Firstly, we introduce a polarized electromagnetic channel model. Secondly, the constraints of dense antenna arrays are analyzed using an infinite array approximation. Thirdly, we conduct numerical simulations and perform a real-world channel measurement experiment to validate the performance boundaries of holographic communications. The main contributions of this paper are summarized as follows:

\begin{itemize}
	\item We propose a polarized electromagnetic channel model for holographic communication. It utilizes the polarization of electromagnetic waves and mutual coupling between antennas, which makes the channel model satisfy the physical constraints.
	\item Based on a theoretical analysis of the radiation characteristics of the dense antenna array, we found that the performance of holographic communication has an inflection point as the array gets denser. In other words, the channel capacity cannot increase boundlessly.
	\item We conduct a channel measurement experiment in an indoor environment. With post-processing, it is demonstrated that element efficiency is the main factor that affects the performance of holographic communication systems. 
\end{itemize}

The rest of this paper is organized as follows. The holographic channel model is briefly reviewed in Section II. Then, the polarized electromagnetic channel model and its relevant concepts are explained in detail in Section III. The related numerical simulations are provided in Section IV. A channel measurement experiment, including the evaluation of spatial oversampling and possible defects in holographic communication, is provided in Section V. Finally, the conclusions and discussions are given in Section VI.

\textit{Notation}: Column vectors and matrices are denoted by lowercase and uppercase boldface letters. The imaginary unit is denoted by $\mathrm{j}$. The modulus operation is denoted by mod($\cdot$,$\cdot$). $\lfloor \cdot \rfloor$ rounds the argument toward negative infinity. Conjugate transposition is denoted by $(\cdot)^\mathrm{H}$. The cardinality of a set is denoted by $|\cdot|$. $\mathbf{I}$ denotes an identity matrix, while $\mathbf{1}$ denotes an all-ones matrix. $[\mathbf{A}]_{ij}$ represents the $i,j$ entry of matrix $\mathbf{A}$. The Hadamard and Kronecker products are $\mathbf{A} \odot \mathbf{B}$ and $\mathbf{A} \otimes \mathbf{B}$, respectively.

\begin{figure*}[!t]
	\centering
	\includegraphics[width=0.9\textwidth]{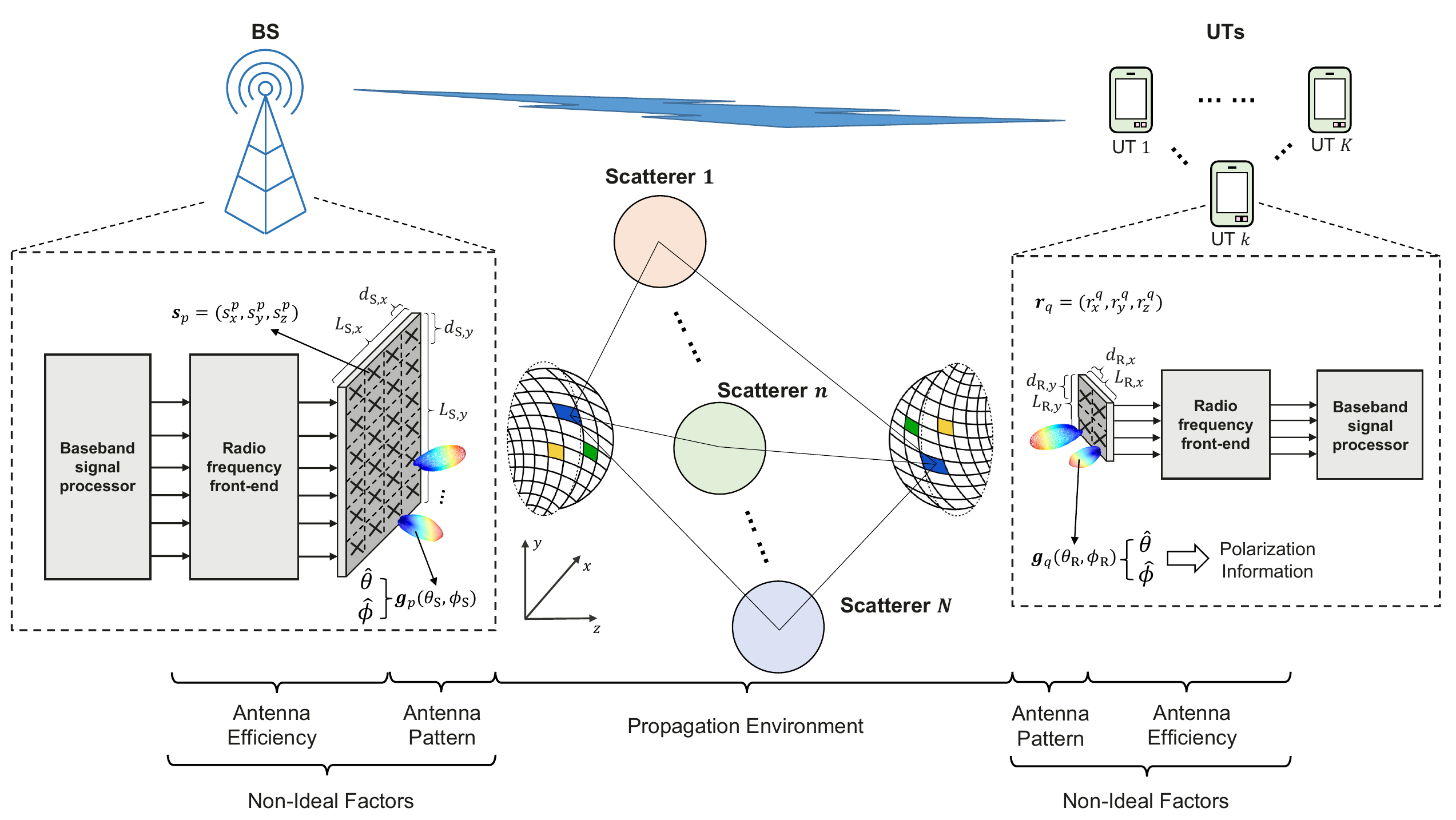}
	\caption{Holographic communication system with homogeneous scattering environment.}
	\label{f1_holo_com}
\end{figure*}

\section{Channel Models in Related Works}

In this section, we briefly review the holographic channel model proposed in \cite{Pizzo2022Fourier}, which is based on the discretization of the Fourier plane-wave series expansion for the electromagnetic field under the NLOS scenario.

Consider the holographic communication system shown in Fig.~\ref{f1_holo_com}. The line-of-sight (LOS) path is blocked, and a large number of homogenous scatterers in the environment provide rich scattering conditions. Both BS and UT employ planar arrays placed in $xy$ plane to transmit or receive radio waves. The width and height of the arrays at the BS side are $L_{\mathrm{S},x}$ and $L_{\mathrm{S},y}$ with antenna spacings $d_{\mathrm{S},x}$, $d_{\mathrm{S},y}$ while at the UT side are $L_{\mathrm{R},x}$ and $L_{\mathrm{R},y}$ with antenna spacings $d_{\mathrm{R},x}$, $d_{\mathrm{R},y}$. Different choice of antenna spacings will result in different antenna quantities $N_\mathrm{S} = N_{\mathrm{S},x} N_{\mathrm{S},y}$ and $N_\mathrm{R} = N_{\mathrm{R},x} N_{\mathrm{R},y}$.

Array elements are indexed row-by-row with $p \in [1,N_\mathrm{S}]$ and $q \in [1,N_\mathrm{R}]$. Local coordinates are established at the center of the array. Therefore, the position of the $p$-th BS array element is
\begin{equation}
	\boldsymbol{s}_{p} = [s_x^p,s_y^p,s_z^p] ^ \mathrm{T} = [u_{\mathrm{S},p} d _{\mathrm{S},x},v_{\mathrm{S},p} d _{\mathrm{S},y},0] ^ \mathrm{T},
\end{equation}
where
\begin{equation}
	\begin{split}
		u_{\mathrm{S},p} &= (-N_{\mathrm{S},x} + \mathrm{mod}(p-1,N_{\mathrm{S},x}))/2, \\
		v_{\mathrm{S},p} &= (N_{\mathrm{S},y} - \lfloor (p-1)/N_{\mathrm{S},y} \rfloor)/2.
	\end{split}
\end{equation}
The position of the $q$-th UT array element is
\begin{equation}
	\boldsymbol{r}_{q} = [r_x^q,r_y^q,r_z^q] ^ \mathrm{T} = [u_{\mathrm{R},q} d _{\mathrm{R},x},v_{\mathrm{R},q} d _{\mathrm{R},y},0] ^ \mathrm{T},
\end{equation}
where
\begin{equation}
	\begin{split}
		u_{\mathrm{R},q} &= (-N_{\mathrm{R},x} + \mathrm{mod}(q-1,N_{\mathrm{R},x}))/2, \\
		v_{\mathrm{R},q} &= (N_{\mathrm{R},y} - \lfloor (q-1)/N_{\mathrm{R},y} \rfloor)/2.
	\end{split}
\end{equation}

\begin{figure}[!t]
	\centering
	\includegraphics[width=0.48\textwidth]{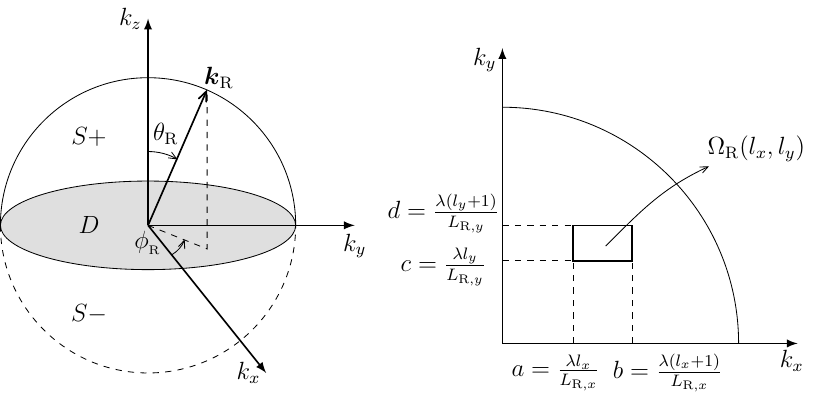}
	\caption{The wavenumber vector of the receive array and its sampling block $\Omega_{\mathrm{R}}(l_x,l_y)$ in the $k_xk_y$ plane.}
	\label{f2_sp_bk}
\end{figure}

In the far-field region, radiation from the source array can be represented as a set of plane waves\cite[Ch.~4]{Prabhakar2022}. Suppose a plane wave propagates outwards from the source array with polar angle $\theta_\mathrm{S}$ and azimuth angle $\phi_\mathrm{S}$, the steering vector of this wave is
\begin{equation}
	\boldsymbol{a}_\mathrm{S}(\theta_\mathrm{S},\phi_\mathrm{S}) = \frac{1}{\sqrt{N_\mathrm{S}}} [e^{- \mathrm{j}\boldsymbol{k}_\mathrm{S}(\theta_\mathrm{S},\phi_\mathrm{S}) \cdot \boldsymbol{s}_1},\cdots,e^{- \mathrm{j} \boldsymbol{k}_\mathrm{S}(\theta_\mathrm{S},\phi_\mathrm{S}) \cdot \boldsymbol{s}_{N_\mathrm{S}}}] ^ \mathrm{T},
	\label{a_s}
\end{equation}
where
\begin{equation}
	\boldsymbol{k}_\mathrm{S}(\theta_\mathrm{S},\phi_\mathrm{S}) = \frac{2\pi}{\lambda}[\mathrm{sin}(\theta_\mathrm{S})\mathrm{cos}(\phi_\mathrm{S}),\mathrm{sin}(\theta_\mathrm{S})\mathrm{sin}(\phi_\mathrm{S}),\mathrm{cos}(\theta_\mathrm{S})]
\end{equation}
is the wavenumber vector at the BS side.

Likewise, we can define the steering vector in the receiver side
\begin{equation}
	\boldsymbol{a}_\mathrm{R}(\theta_\mathrm{R},\phi_\mathrm{R}) = \frac{1}{\sqrt{N_\mathrm{R}}} [e^{- \mathrm{j} \boldsymbol{k}_\mathrm{R}(\theta_\mathrm{R},\phi_\mathrm{R}) \cdot \boldsymbol{r}_1},\cdots,e^{- \mathrm{j} \boldsymbol{k}_\mathrm{R}(\theta_\mathrm{R},\phi_\mathrm{R}) \cdot \boldsymbol{r}_{N_\mathrm{R}}}] ^ \mathrm{T},
	\label{a_r}
\end{equation}
where
\begin{equation}
	\boldsymbol{k}_\mathrm{R}(\theta_\mathrm{R},\phi_\mathrm{R}) = \frac{2\pi}{\lambda}[\mathrm{sin}(\theta_\mathrm{R})\mathrm{cos}(\phi_\mathrm{R}),\mathrm{sin}(\theta_\mathrm{R})\mathrm{sin}(\phi_\mathrm{R}),\mathrm{cos}(\theta_\mathrm{R})]
\end{equation}
is the wavenumber vector at the UT side.

When the array size is large, it has a higher angular resolution in the wavenumber domain, leading to the possibility of discretization of the angular response. Following the Nyquist sampling theorem, a planar array with size ($L_x,L_y$) should have a maximum sampling interval ($2\pi /L_x, 2\pi /L_y$) in the wavenumber domain, as shown in Fig.~\ref{f2_sp_bk}. Denote the sampling index by ($m_x,m_y$) in $\boldsymbol{k}_\mathrm{S}$ plane and ($l_x,l_y$) in $\boldsymbol{k}_\mathrm{R}$ plane. When $k_x^2 + k_y^2 > k_0^2$, $k_z$ is pure imaginary and corresponds to evanescent waves \cite{Prabhakar2022}. Consequently, the consideration is narrowed down to indices $(m_x,m_y) \in \mathcal{E}_\mathrm{S}$ and $(l_x,l_y) \in \mathcal{E}_\mathrm{R}$. These indices constitute finite sets shaping the semi-sphere within the wavenumber domain, as shown in Fig.~\ref{f2_sp_bk}.

Since each sampling index pair $\boldsymbol{m} = (m_x,m_y)$ or $\boldsymbol{l} = (l_x,l_y)$ corresponds to an angle pair $(\theta,\phi)$, $\mathbf{a}_\mathrm{S}(\theta_\mathrm{S},\phi_\mathrm{S})$ and $\mathbf{a}_\mathrm{R}(\theta_\mathrm{R},\phi_\mathrm{R})$ can be represented by $\mathbf{a}_\mathrm{S}(\boldsymbol{m})$ and $\mathbf{a}_\mathrm{R}(\boldsymbol{l})$. Based on the discrete Fourier plane-wave series expansion, the holographic channel matrix can be written as \cite{Pizzo2022Fourier}
\begin{equation}
	\mathbf{H} = \sqrt{N_{\mathrm{R}}N_{\mathrm{S}}} \times \sum_{\boldsymbol{l} \in \mathcal{E}_\mathrm{R}}{\sum_{\boldsymbol{m} \in \mathcal{E}_\mathrm{S}}{H_a\left( \boldsymbol{l},\boldsymbol{m} \right)} \boldsymbol{a}_\mathrm{R}(\boldsymbol{l}) \boldsymbol{a}_\mathrm{S}^{\mathrm{T}}(\boldsymbol{m})},
	\label{H_holo}
\end{equation}
where the small-scale fading is modeled as a zero-mean, spatially-stationary, and correlated Gaussian random field, $H_a\left( \boldsymbol{l},\boldsymbol{m} \right) \sim \mathcal{N} _{\mathbb{C}}\left( 0,\sigma ^2(\boldsymbol{l},\boldsymbol{m}) \right)$ is the angular response that maps the $\boldsymbol{m}$-th outgoing wave to the $\boldsymbol{l}$-th incident wave with the variance $\sigma ^2(\boldsymbol{l},\boldsymbol{m})$ given by \cite{Pizzo2020Spatially}
\begin{equation}
	\begin{split}
		\sigma ^2(\boldsymbol{l},\boldsymbol{m}) = &\iint_{\Omega _{\mathrm{R}}\left( l_x,l_y \right)}{A_{\mathrm{R}}^{2}\left( \theta _{\mathrm{R}},\phi _{\mathrm{R}} \right)}\sin \theta _{\mathrm{R}}\mathrm{d}\theta _{\mathrm{R}}\mathrm{d}\phi _{\mathrm{R}} \times \\
		&\iint_{\Omega _{\mathrm{S}}\left( m_x,m_y \right)}{A_{\mathrm{S}}^{2}\left( \theta _{\mathrm{S}},\phi _{\mathrm{S}} \right)}\sin \theta _{\mathrm{S}}\mathrm{d}\theta _{\mathrm{S}}\mathrm{d}\phi _{\mathrm{S}},
	\end{split}
\end{equation}
where $\Omega _{\mathrm{R}}$, $\Omega _{\mathrm{S}}$ are the integration region that correspond to the sampling block in the wavenumber domain \cite{Pizzo2022Fourier}, the spectral factors $A_\mathrm{R}(\theta,\phi)$ and $A_\mathrm{S}(\theta,\phi)$ denote the impact of the propagation environment in the wavenumber domain. For simplicity, in this paper we only consider the isotropic environment. In this case, $A_\mathrm{R}(\theta,\phi) = A_\mathrm{S}(\theta,\phi) = 1$.

The holographic channel model can also be written in matrix form \cite{Pizzo2022Fourier}:
\begin{equation}
	\mathbf{H}_{\mathrm{holo}} = \mathbf{U}_\mathrm{R} \mathbf{H}_\mathrm{a} \mathbf{U}_\mathrm{S}^\mathrm{T},
	\label{H_holo_matr}
\end{equation}
where $\mathbf{H}_\mathrm{a} \in \mathbb{C}^{n_\mathrm{R} \times n_\mathrm{S}}$ collects the angular responses between transceiver arrays in the wavenumber domain, $\mathbf{U}_\mathrm{R} \in \mathbb{C}^{N_\mathrm{R} \times n_\mathrm{R}}$ and $\mathbf{U}_\mathrm{S} \in \mathbb{C}^{N_\mathrm{S} \times n_\mathrm{S}}$ are the steering vectors of the receive array and the source array, $n_\mathrm{S} = |\mathcal{E}_\mathrm{S}|$ and $n_\mathrm{R} = |\mathcal{E}_\mathrm{R}|$ are the number of angle samples in the source array and the receive array, respectively. However, two important factors of the wireless channel are not included in this model. One is the polarization of electromagnetic waves; the other is the mutual coupling among array elements. In Eq.~\eqref{H_holo_matr}, BS and UTs communicate without polarization, that is, via monochromatic scalar waves. In addition, it is assumed that all array elements are ideal isotropic antennas. Therefore, we should take these realistic factors into account to build a more accurate and practical channel model.

\section{Holographic Channel Model with Physical Constraints}

In this section, we analyze the effect of mutual coupling from three perspectives: embedded element pattern, active impedance, and total efficiency of an array element. Leveraging insights from these perspectives, we incorporate mutual coupling considerations to propose a more realistic channel model—the polarized electromagnetic channel model, which also takes into account the polarization of electromagnetic waves. Based on this modified model, we found that closely spaced elements in a limited space cannot increase the channel performance boundlessly.

\subsection{Embedded Element Pattern and Active Impedance}

When an antenna is deployed within an array, particularly a dense antenna array, mutual coupling among elements significantly influences its radiation characteristics. Embedded element pattern and active impedance are two terms that illustrate the effect of mutual coupling.

If we excite a single antenna inside an array, the electromagnetic waves generated by this element will induce currents on other elements. As a result, these elements also radiate electromagnetic waves, which contribute to the overall radiation along with the waves from the initially excited element, thereby causing pattern distortion. Therefore, it is necessary to define the antenna pattern in an array, which is called the embedded element pattern \cite{Pozar1994}.

\begin{figure}[!t]
	\centering
	\includegraphics[width=0.45\textwidth]{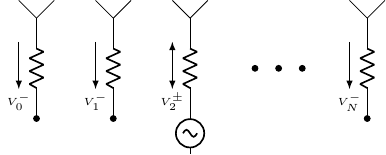}
	\caption{An $N$-element uniform linear array excited by a voltage source.}
	\label{f3_act_ele_patt}
\end{figure}

Consider the linear array shown in Fig.~\ref{f3_act_ele_patt}. All elements, except for one connected to a voltage generator, are terminated with matched loads. The induced currents on the non-excited elements also radiate electromagnetic waves, which change the radiation characteristics of the excited element. Additionally, if the non-excited elements possess different termination statuses (e.g., short-circuited or open-circuited), the pattern of the excited antenna behaves differently. Hence, the embedded element pattern is directly related to the termination status of all antenna elements. Denote the embedded element gain pattern of the $n$-th antenna by $\boldsymbol{g}_{n}(\theta,\phi)$, where non-excited elements are terminated with matched loads. When all the elements are driven by voltage sources, the gain of the array can be represented as \cite{Pozar1994}
\begin{equation}
	\boldsymbol{g}_{\mathrm{array}}(\theta,\phi) = \sum_{n=1}^{N} V_n \boldsymbol{g}_{n}(\theta,\phi),
\end{equation}
where $V_n$ is the source voltage of the $n$-th element.

The relationship between the embedded element directivity pattern $\boldsymbol{d}_n(\theta,\phi)$ and $\boldsymbol{g}_n(\theta,\phi)$ is given by
\begin{equation}
	\boldsymbol{g}_n(\theta,\phi) = \sqrt{\chi_{_{0,n}}} \boldsymbol{d}_n(\theta,\phi),
\end{equation}
where $\chi_{_{0,n}}$ is the antenna efficiency of the $n$-th antenna element in the array. The square root is due to the relationship between power pattern and field pattern.

Active impedance is closely related to antenna efficiency. By adjusting the amplitude and phase of each element's driven signal, an array can steer its beam to a desired angle, namely the scan angle. If all the elements are linearly phased with uniform amplitudes, this arrangement of source is named Floquet excitation \cite{Arun2005}, which allows us to analyze the characteristics of central elements inside a phased array using the results deduced from an infinite array.

Once the structure of the array is fixed, we can derive its scattering matrix $\mathbf{S}$ through electromagnetic simulation or measurement. Notice that $[\mathbf{S}]_{ij}$ is defined as the ratio of the reflected wave $V_i^-$ measured on port $i$ to the incident wave $V_j^+$ measured on port $j$, that is \cite{Pozar2012},
\begin{equation}
	[\mathbf{S}]_{ij} = \left. \frac{V_i^-}{V_j^+} \right|_{V_{k}^{+}=0~\mathrm{for}~k\ne j},
\end{equation}
where $V_k^+ = 0$ means that these ports are terminated with matched load.

The impedance matrix $\mathbf{Z}$ is another measurable quantity for mutual coupling. Similarly, $[\mathbf{Z}]_{ij}$ is defined as the ratio of the total voltage $V_i$ measured on port $i$ to the total current $I_j$ measured on port $j$ \cite{Pozar2012}:
\begin{equation}
	[\mathbf{Z}]_{ij} = \left. \frac{V_i}{I_j} \right|_{I_{k}=0~\mathrm{for}~k\ne j},
\end{equation}
where $I_{k}=0$ means that these ports are terminated with open-circuit.

The matrices $\mathbf{S}$ and $\mathbf{Z}$ are related to each other as follows:
\begin{equation}
	\mathbf{Z} = Z_0 (\mathbf{I} + \mathbf{S}) (\mathbf{I} - \mathbf{S})^{-1},
\end{equation}
where $Z_0$ is the characteristic impedance of the transmission line, in most cases, $Z_0 = 50~\Omega$.

The input impedance of the $m$-th array element is defined as
\begin{equation}
	Z_{\mathrm{in},m} = \frac{V_m}{I_m} = \sum_{n=1}^N{[\mathbf{Z}]_{mn}\frac{I_n}{I_m}}.
	\label{Z_in}
\end{equation}
Notice that $Z_{\mathrm{in},m}$ is related to the current of each element, which changes as the array scans at different angles. As a consequence, $Z_{\mathrm{in},m}$ varies according to \eqref{Z_in}. Therefore, the input impedance of an array element is named active impedance, which means that it varies with the excitation.

During the measurement of the embedded element pattern, the active impedance of the array element is different from its input impedance when it is isolated. Therefore, the matching network of array elements should be carefully designed, especially for dense arrays. In an infinite regular array, the active impedance of each array element is the same, which leads to the possibility of calculating the theoretical upper bound of the antenna efficiency. For a large-scale array, the scattering environments of its central elements are almost the same, and thus they have similar radiation characteristics, including antenna efficiency and radiation pattern \cite{Craeye2011}.

\subsection{Antenna Efficiency}

When an antenna is isolated, its total efficiency $\chi_{_{0}}$ is defined as the ratio of the radiation power to the input power delivered to its transmission lines \cite{Balanis2016}. Specifically, it can be represented as the product of the radiation efficiency $\chi_\mathrm{r}$ and the transmission efficiency $\chi_\mathrm{t}$:
\begin{equation}
	\chi_{_0} = \chi_\mathrm{r} \chi_\mathrm{t}.
\end{equation}
In the following section, we analyze the characteristics of $\chi_\mathrm{r}$ and $\chi_\mathrm{t}$, to find the upper bound of its total efficiency $\chi_{_0}$.

\subsubsection{Radiation Efficiency}
The total power delivered to an antenna is consumed by two parts: one is the radiation power $P_{\mathrm{rad}}$; the other is the conductor loss $P_{\mathrm{loss}}$. Radiation efficiency $\chi_\mathrm{r}$ of an antenna is defined as the ratio of the radiation power to the total power delivered to the antenna, that is,
\begin{equation} \label{eff_r}
	\chi_\mathrm{r} = \frac{P_{\mathrm{rad}}}{P_{\mathrm{rad}} + P_{\mathrm{loss}}} = \frac{1}{1 + P_{\mathrm{loss}}/P_{\mathrm{rad}}}.
\end{equation}

The power radiated by a current source $\mathbf{J}$ with frequency $\omega$ inside the region $\mathcal{V}$ can be computed by \cite{Vandenbosch2010}
\begin{equation}
	\begin{split}
		P_\mathrm{rad} = &\frac{k_0}{8\pi\omega\epsilon_0} \int_{\mathcal{V}_1} \int_{\mathcal{V}_2} \{k_0^2[\mathbf{J(r_1)} \cdot \mathbf{J^\ast(r_2)}]-\\
		&[\nabla_1\cdot\mathbf{J(r_1)}] [\nabla_2\cdot\mathbf{J^\ast(r_2)}]\} \frac{\sin(k_0 R)}{k_0 R} \mathrm{d} \mathcal{V}_1 \mathrm{d} \mathcal{V}_2,
	\end{split}
\end{equation}
where $k_0$ is the wavenumber, $\epsilon_0$ is the permittivity constant in free space, $R=|\mathbf{r}_1-\mathbf{r}_2|$, $\mathrm{d} \mathcal{V}_1$ and $\mathrm{d} \mathcal{V}_2$ are the volume unit near $\mathbf{r}_1$ and $\mathbf{r}_2$. Expand $\sin(k_0R)/(k_0R)$ with Taylor series and take the first two items, that is,
\begin{equation}
	\frac{\sin(k_0 R)}{k_0 R} \approx 1 - \frac{(kR)^2}{6},
\end{equation}
we obtain \cite{Shahpari2018}
\begin{equation} \label{P_rad}
	P_\mathrm{rad} = \frac{k_0^2\eta_0}{12\pi} \left|\iiint_\mathcal{V} \mathbf{J}~\mathrm{d} \mathcal{V} \right|^2 \le \frac{k_0^2\eta_0}{12\pi} \frac{S}{\alpha^2} \iint_S |\mathbf{J}_\mathrm{s}|^2 ~\mathrm{d} S,
\end{equation}
where $S$ is the surface of $\mathcal{V}$, $\mathbf{J}_\mathrm{s}$ is the surface current, $\eta_0$ is the wave impedance of free space, and $\alpha=\sqrt{\pi f\mu\sigma}$, whose inverse $1/\alpha$ is the skin depth $\delta$.

\begin{figure}[!t]\centering
	\includegraphics[width=0.45\textwidth]{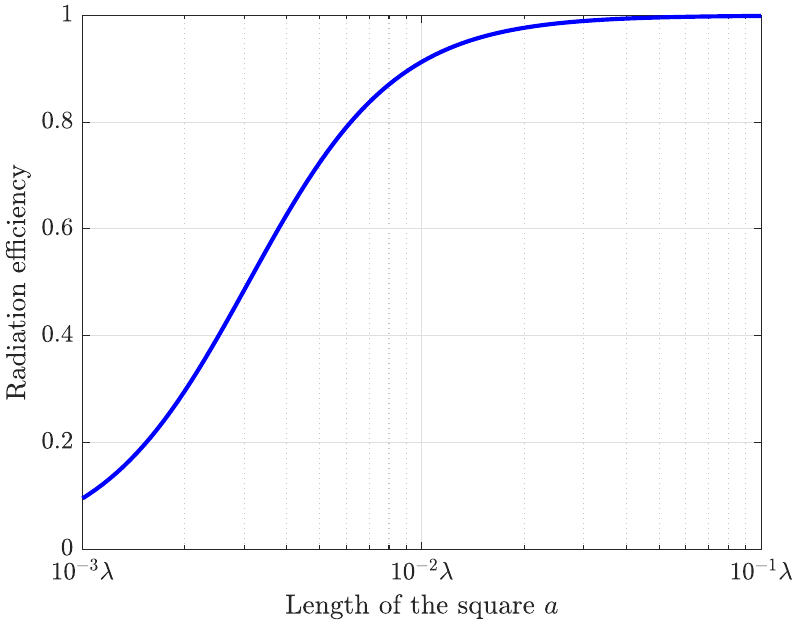}
	\caption{The upper bound of radiation efficiency with different antenna sizes.}
	\label{f4_rad_eff}
\end{figure}

According to Joule's law, the conductor loss of an antenna is
\begin{equation} \label{P_loss}
	P_\mathrm{loss} = \frac{1}{2\sigma} \iiint_\mathcal{V} |\mathbf{J}|^2 ~\mathrm{d} \mathcal{V} = \frac{1}{4\sigma\alpha} \iint_S |\mathbf{J}_\mathrm{s}|^2 ~\mathrm{d} S.
\end{equation}
Eliminating $P_\mathrm{rad}$ and $P_\mathrm{loss}$ from \eqref{P_rad} and \eqref{P_loss} in \eqref{eff_r}, we can finally obtain \cite{Shahpari2018}
\begin{equation}
	\chi_\mathrm{r} = \frac{1}{1 + P_\mathrm{loss}/P_\mathrm{rad}} \le \left(1 + \frac{3\pi}{2}\frac{\delta}{kS}\right)^{-1}.
\end{equation}

Suppose we have a square antenna with side length $a$, which is made from aluminium with conductivity $\sigma=3.5\times 10^7$ S/m. Its operating frequency is at 2 GHz, resulting in a skin depth $\delta=1.9\times 10^{-6}$ m. The relation between $\chi_\mathrm{r}$ and $a$ is shown in Fig.~\ref{f4_rad_eff}. We can find that for this antenna, when $a=\lambda/10$, $\chi_\mathrm{r} \approx 1$, and when $a=\lambda/100$, $\chi_\mathrm{r}\approx 0.9$. Since most antennas have side lengths larger than $\lambda/10$, the power consumed by conductor loss is negligible, and we assume $\chi_\mathrm{r} = 1$ in the following analysis. In this case, the total efficiency $\chi_{_0}$ of an array antenna is mainly determined by its transmission efficiency $\chi_\mathrm{t}$.

\subsubsection{Transmission Efficiency}
The efficiency $\chi_\mathrm{t}$ of an array antenna is the ratio of the radiation power of the array to the incident power of its connected transmission line, when only that element is excited,
\begin{equation} \label{eff_t}
	\chi_{\mathrm{t}} = \frac{P_\mathrm{rad}}{P_\mathrm{inc}} = 1 - \frac{P_\mathrm{ref}}{P_\mathrm{inc}},
\end{equation}
where $P_\mathrm{ref}$ is the power reflected back into the array elements, including the excited element and other elements. If the element is driven by a source with unit amplitude, the scattering parameters $S_{p,q}$ can represent the reflected voltage. In this case, the transmission efficiency of the array antenna can be calculated by \cite{Hannan1964}
\begin{equation}
	\chi_{\mathrm{t}} = 1 - \frac{P_\mathrm{ref}}{P_\mathrm{inc}} = 1 - \sum \nolimits_{p,q} |S_{p,q}|^2.
	\label{S_chi_relation}
\end{equation}

\begin{figure}[!t]\centering
	\includegraphics[width=0.45\textwidth]{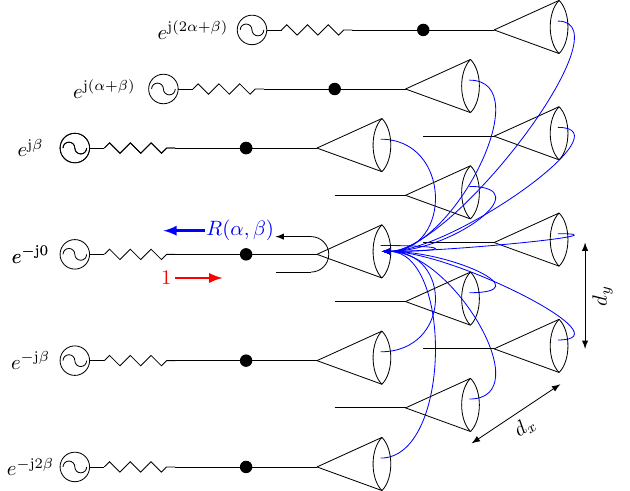}
	\caption{An infinite planar array with rectangular grids.}
	\label{f5_act_ref}
\end{figure}

For a finite array, elements are subject to mutual coupling at different levels, making it challenging to establish a universal efficiency limit for each element. However, if we consider an infinite array, {where} each element encounters the same scattering environment, it will be more tractable to obtain the upper bound for $\chi_{\mathrm{t}}$.

The active reflection coefficient $R(\theta,\phi)$ of an array element is defined as the reflection coefficient at the terminals of an array element when all array elements are excited. Similar to the active impedance, it is also a function of the scan angle $(\theta,\phi)$. In an infinite planar array with rectangular grids shown in Fig.~\ref{f5_act_ref}, the active reflection coefficient $R(\theta,\phi)$ can be computed from the scattering parameters $S_{p,q}$ as follows \cite{Hannan1964}:
\begin{equation} \label{ac_ref}
	R(\psi_x,\psi_y) = \sum_{p=-\infty}^\infty\sum_{q=-\infty}^\infty S_{p,q} \exp(-\mathrm{j}p\psi_x-\mathrm{j}q\psi_y).
\end{equation}
Note that the variables $(\theta,\phi)$ have been replaced by $(\psi_x,\psi_y)$, where
\begin{equation}
	\psi_x = kd_x\sin(\theta)\cos(\phi),~\psi_y = kd_y\sin(\theta)\sin(\phi).
\end{equation}
According to Parseval's theorem, it follows that \cite{Kahn1967}
\begin{equation}
	\sum_{p=-\infty}^{\infty}\sum_{q=-\infty}^{\infty} |S_{p,q}|^2 = \frac{1}{4\pi^2}\int_{-\pi}^{\pi}\int_{-\pi}^{\pi} |R(\psi_x,\psi_y)|^2 ~ \mathrm{d} \psi_x \mathrm{d} \psi_y.
	\label{S_R_relation}
\end{equation}
Therefore, the transmission efficiency $\chi_{\mathrm{t}}$ can be determined by the active reflection coefficient $R(\psi_x,\psi_y)$:
\begin{equation} \label{eff_e}
	\begin{split}
		\chi_{\mathrm{t}} &= 1 - \sum_{p=-\infty}^{\infty}\sum_{q=-\infty}^{\infty} |S_{p,q}|^2 \\
		&= 1 - \frac{1}{4\pi^2}\int_{-\pi}^{\pi}\int_{-\pi}^{\pi} |R(\psi_x,\psi_y)|^2 ~\mathrm{d}\psi_x\mathrm{d}\psi_y.
	\end{split}
\end{equation}

For an infinite array, its array factor tends to be a Dirac delta pseudo-function, and the energy is concentrated on the desired beam. Thus, we can take the following approximation in the calculation of \eqref{eff_e} \cite{Hannan1964,Kahn1967}:
\begin{enumerate}
	\item In the visible region, array elements are totally matched, $R(\psi_x,\psi_y) = 0$.
	\item In the invisible region, an array cannot radiate outwards, $R(\psi_x,\psi_y) = 1$.
\end{enumerate}

\begin{figure}[!t]
	\centering
	\includegraphics[width=0.45\textwidth]{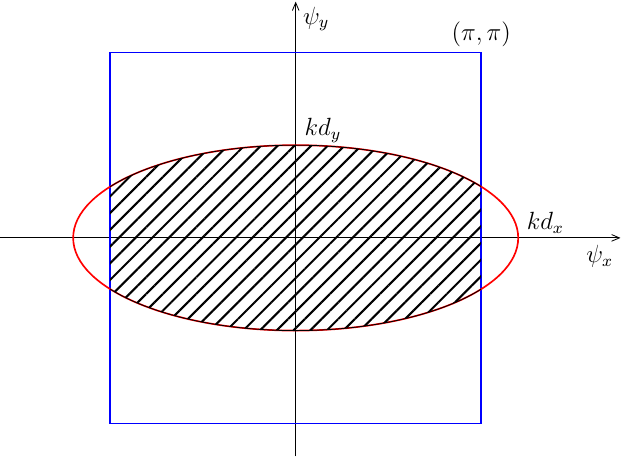}
	\caption{The visible region (ellipse) of a rectangular grid array with $d_x = 0.6\lambda$ and $d_y = 0.25\lambda$. In the intersection area, $R(\psi_x,\psi_y)$ is assumed to be 0; and in the region outside ellipse but inside the square, $R(\psi_x,\psi_y)$ is assumed to be 1.}
	\label{f6_crt_pt}
\end{figure}

\begin{figure}[!t]
	\centering
	\includegraphics[width=0.45\textwidth]{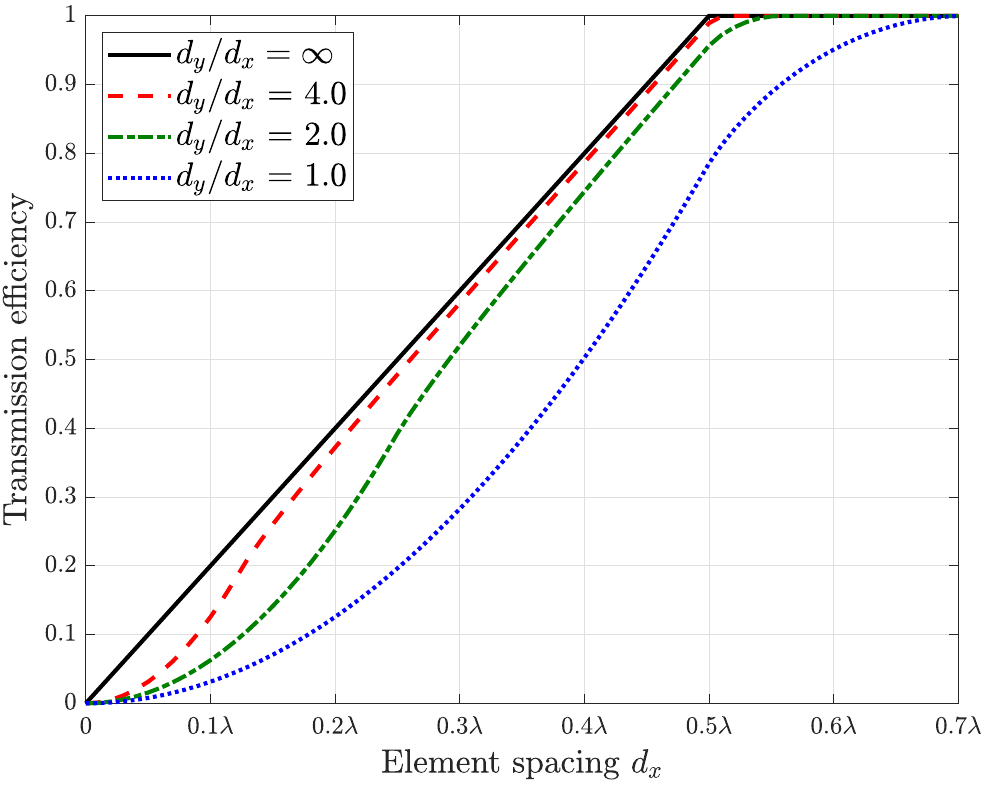}
	\caption{The theoretical upper bound of transmission efficiency for an infinite planar array with rectangular grids.}
	\label{f7_ele_eff}
\end{figure}

The visible region of an array corresponds to the physically observable interval $(\psi_x,\psi_y)$ in the array factor, and is related to the element spacings $d_x$ and $d_y$. Fig.~\ref{f6_crt_pt} shows the relationship between the visible region (the area bounded by the ellipse) and the integral region (the area bounded by the square) when $d_x = 0.6\lambda$ and $d_y = 0.25\lambda$. For a rectangular infinite array, the upper bound of $\chi_\mathrm{t}$ for different antenna spacings is shown in Fig.~\ref{f7_ele_eff}. We can observe that the upper bound decreases rapidly as the antenna spacing gets closer. The limit for different grid structures can also be obtained in a similar way \cite{Kahn1967}. This efficiency upper bound has been verified in \cite{Kildal2016} (see also Sec. IV of this paper).

\subsection{Polarized Electromagnetic Channel Model}

The electromagnetic waves have polarization information describing the orientation of the oscillating electric field. At a specific angle pair $(\theta,\phi)$, the directivity components in $\hat{\boldsymbol{\theta}}$ direction and $\hat{\boldsymbol{\phi}}$ direction are given by $d_\theta(\theta,\phi)$ and $d_\phi(\theta,\phi)$, respectively. Ideally, the cross-polar transmissions between a $\hat{\boldsymbol{\theta}}$ polarized transmit antenna to a $\hat{\boldsymbol{\phi}}$ polarized receive antenna is zero \cite{Oestges2008}. However, in practice, this conclusion is not valid due to the correlation between the orthogonally polarized propagation channels. Cross-polar discrimination (XPD) is defined as the ratio of the co-polarization component to the cross-polarization component, which can be further divided into two factors: the cross-polar isolation (XPI) within  antennas and the cross-polar ratio (XPR) caused by the propagation environment. The effect of XPI has been coupled with the distortion of the embedded element patterns $d_\theta(\theta,\phi)$ and $d_\phi(\theta,\phi)$. As a consequence, we only need to consider the factor of XPR, and model them together in the polarized electromagnetic channel model.

\begin{figure*}[!t]
	\normalsize
	\setcounter{mytempeqncnt}{\value{equation}}
	\setcounter{equation}{35}
	\begin{subequations}
		\label{F_mtx}
		\begin{align}
			\mathbf{F}_{\mathrm{R}} &= 
			\left[ \begin{matrix}	\begin{matrix}	d_{\mathrm{R},\theta ,1}\left( \theta _1,\phi _1 \right)&		d_{\mathrm{R},\phi ,1}\left( \theta _1,\phi _1 \right)\\\end{matrix}&		\cdots&		\begin{matrix}	d_{\mathrm{R},\theta ,1}\left( \theta _{n_{\mathrm{R}}},\phi _{n_{\mathrm{R}}} \right)&		d_{\mathrm{R},\phi ,1}\left( \theta _{n_{\mathrm{R}}},\phi _{n_{\mathrm{R}}} \right)\\\end{matrix}\\	\vdots&		\ddots&		\vdots\\	\begin{matrix}	d_{\mathrm{R},\theta ,N_{\mathrm{R}}}\left( \theta _1,\phi _1 \right)&		d_{\mathrm{R},\phi ,N_{\mathrm{R}}}\left( \theta _1,\phi _1 \right)\\\end{matrix}&		\cdots&		\begin{matrix}	d_{\mathrm{R},\theta ,N_{\mathrm{R}}}\left( \theta _{n_{\mathrm{R}}},\phi _{n_{\mathrm{R}}} \right)&		d_{\mathrm{R},\phi ,N_{\mathrm{R}}}\left( \theta _{n_{\mathrm{R}}},\phi _{n_{\mathrm{R}}} \right)\\\end{matrix}\\\end{matrix} \right]\\
			\mathbf{F}_{\mathrm{S}} &= 
			\left[ \begin{matrix}	\begin{matrix}	d_{\mathrm{S},\theta ,1}\left( \theta _1,\phi _1 \right)&		d_{\mathrm{S},\phi ,1}\left( \theta _1,\phi _1 \right)\\\end{matrix}&		\cdots&		\begin{matrix}	d_{\mathrm{S},\theta ,1}\left( \theta _{n_{\mathrm{S}}},\phi _{n_{\mathrm{S}}} \right)&		d_{\mathrm{S},\phi ,1}\left( \theta _{n_{\mathrm{S}}},\phi _{n_{\mathrm{S}}} \right)\\\end{matrix}\\	\vdots&		\ddots&		\vdots\\	\begin{matrix}	d_{\mathrm{S},\theta ,N_{\mathrm{S}}}\left( \theta _1,\phi _1 \right)&		d_{\mathrm{S},\phi ,N_{\mathrm{S}}}\left( \theta _1,\phi _1 \right)\\\end{matrix}&		\cdots&		\begin{matrix}	d_{\mathrm{S},\theta ,N_{\mathrm{S}}}\left( \theta _{n_{\mathrm{S}}},\phi _{n_{\mathrm{S}}} \right)&		d_{\mathrm{S},\phi ,N_{\mathrm{S}}}\left( \theta _{n_{\mathrm{S}}},\phi _{n_{\mathrm{S}}} \right)\\\end{matrix}\\\end{matrix} \right]
		\end{align}
	\end{subequations}
	\setcounter{equation}{\value{mytempeqncnt}}
	\hrulefill
	\vspace*{4pt}
\end{figure*}

The effect of the propagation environment on depolarization can be separated into two parts: the fading of amplitude and the change of phase. Both of them can be modeled as random variables that take different realizations in different clusters of electromagnetic waves \cite{3GPP38901}, that is, different sampling points in the wavenumber domain. The entry of the polarized channel model can be obtained as
\begin{equation}
	\begin{split}
		[\mathbf{H}]_{qp} \approx &\sum_{\boldsymbol{l} \in \mathcal{E}_\mathrm{R}}{\sum_{\boldsymbol{m} \in \mathcal{E}_\mathrm{S}}{ \sqrt{\chi_{_{0,\mathrm{S},p}} \chi_{_{0,\mathrm{R},q}}}
				\left[ \begin{array}{c}	d_{\mathrm{R},\theta,q}\left( \boldsymbol{l} \right)\\	d_{\mathrm{R},\phi,q}\left( \boldsymbol{l} \right)\\\end{array} \right] ^\mathrm{T} }}\\
		&\left( \mathbf{P}_{\boldsymbol{l},\boldsymbol{m}} \cdot H_\mathrm{a}\left( \boldsymbol{l},\boldsymbol{m} \right) \right) \left[ \begin{array}{c}	d_{\mathrm{S},\theta,p}\left( \boldsymbol{m} \right)\\	d_{\mathrm{S},\phi,p}\left( \boldsymbol{m} \right)\\\end{array} \right].
	\end{split}
	\label{ch4_H_single}
\end{equation}
At the specified angle, $d_{\mathrm{R},\theta,q}\left( \boldsymbol{l} \right)$ and $d_{\mathrm{R},\phi,q}\left( \boldsymbol{l} \right)$ are the $\hat{\boldsymbol{\theta}}$ and $\hat{\boldsymbol{\phi}}$ components of the embedded element directivity pattern of the $q$-th antenna in the receive array, while $d_{\mathrm{S},\theta,p}\left( \boldsymbol{m} \right)$ and $d_{\mathrm{S},\phi,p}\left( \boldsymbol{m} \right)$ are the $\hat{\boldsymbol{\theta}}$ and $\hat{\boldsymbol{\phi}}$ components of the embedded element directivity pattern of the $p$-th antenna in the source array.
\begin{equation}
	\mathbf{P}_{\boldsymbol{l},\boldsymbol{m}} = \frac{1}{\sqrt{1+\kappa _{\boldsymbol{l},\boldsymbol{m}}^{-1}}} \left[ \begin{matrix}	e^{\mathrm{j} \Phi _{\boldsymbol{l},\boldsymbol{m}}^{\theta \theta}}&		\sqrt{\kappa _{\boldsymbol{l},\boldsymbol{m}}^{-1}} e^{\mathrm{j}\Phi _{\boldsymbol{l},\boldsymbol{m}}^{\theta \phi}}\\	\sqrt{\kappa _{\boldsymbol{l},\boldsymbol{m}}^{-1}} e^{\mathrm{j} \Phi _{\boldsymbol{l},\boldsymbol{m}}^{\phi \theta}}&	e^{\mathrm{j} \Phi _{\boldsymbol{l},\boldsymbol{m}}^{\phi \phi}}\\\end{matrix} \right]
\end{equation}
denotes the polarization leakage of the propagation path determined by $\boldsymbol{l}$ and $\boldsymbol{m}$, where $\Phi _{\boldsymbol{l}, \boldsymbol{m}}^{\theta \theta}$, $\Phi _{\boldsymbol{l},\boldsymbol{m}}^{\theta \phi}$, $\Phi _{\boldsymbol{l},\boldsymbol{m}}^{\phi \phi}$, $\Phi _{\boldsymbol{l},\boldsymbol{m}}^{\phi \theta}$ are the random phase shifts that follow the uniform distribution between $0$ and $2\pi$, and $\kappa _{\boldsymbol{l},\boldsymbol{m}} = 10^{X_{\boldsymbol{l},\boldsymbol{m}}/10}$ is the cross-polarization power ratio, in which $X_{\boldsymbol{l},\boldsymbol{m}}$ follows the Gaussian distribution with mean value $\mu_{\mathrm{_{XPR}}}$ and standard deviation $\sigma_{_\mathrm{XPR}}$. The typical values of $\mu_{\mathrm{_{XPR}}}$ and $\sigma_{_\mathrm{XPR}}$ can be found in the standard of 3GPP TR 38.901 \cite{3GPP38901}.

After obtaining each entry of the channel transfer matrix, we can further write the complete electromagnetic channel model in matrix form
\begin{equation}
	\mathbf{H}_{\mathrm{pol}} = \mathbf{\Gamma}_\mathrm{R}\mathbf{F}_\mathrm{R}\mathbf{\Omega }\odot \left( \mathbf{H}_\mathrm{a}\otimes \mathbf{1}_{2\times 2} \right) \mathbf{F}_\mathrm{S}^{\mathrm{T}} \mathbf{\Gamma}_\mathrm{S}^\mathrm{T},
	\label{H_em_plr}
\end{equation}
where
\begin{equation}
	\mathbf{\Omega }=\left[ \begin{matrix}	\mathbf{P}_{\boldsymbol{l}_1,\boldsymbol{m}_1}&		\cdots&		\mathbf{P}_{\boldsymbol{l}_1,\boldsymbol{m}_{n_\mathrm{S}}}\\	\vdots&		\ddots&		\vdots\\	\mathbf{P}_{\boldsymbol{l}_{n_\mathrm{R}},\boldsymbol{m}_1}&		\cdots&		\mathbf{P}_{\boldsymbol{l}_{n_\mathrm{R}},\boldsymbol{m}_{n_\mathrm{S}}}\\\end{matrix} \right]
\end{equation}
collects the polarization leakage of all propagation paths.
\begin{subequations}
	\begin{align}
		\mathbf{\Gamma }_{\mathrm{R}} &= \text{diag}(\sqrt{\chi_{_{0,\mathrm{R},1}}}, \cdots, \sqrt{\chi_{_{0,\mathrm{R},N_\mathrm{R}}}}), \\
		\mathbf{\Gamma }_{\mathrm{S}} &= \text{diag}(\sqrt{\chi_{_{0,\mathrm{S},1}}}, \cdots, \sqrt{\chi_{_{0,\mathrm{S},N_\mathrm{S}}}})
	\end{align}
\end{subequations}
are two diagonal matrices that collect the square root of the antenna efficiency. $\mathbf{H}_\mathrm{a}\in \mathbb{C}^{n_\mathrm{R} \times n_\mathrm{S}}$ is derived from \eqref{H_holo_matr}. $\mathbf{F}_\mathrm{R}$ and $\mathbf{F}_\mathrm{S}$, given in \eqref{F_mtx}, collect the embedded element directivity patterns, including the $\hat{\boldsymbol{\theta}}$ component and $\hat{\boldsymbol{\phi}}$ component.

\section{Numerical Simulations}

In this section, we use isotropic antennas and half-wavelength dipoles to perform numerical simulations. The performance of the holographic system is evaluated by its channel capacity.

\subsection{Array of Isotropic Antennas}

Theoretically, an isotropic antenna is a point source whose radiation intensity is a constant, and thus its directivity is 0 dBi. However, this is not the case if we consider the element efficiency loss brought by dense array arrangements. Suppose two isotropic antenna arrays communicate in the form of scalar waves. If we ignore the antenna pattern distortion owing to mutual coupling, the channel matrix \eqref{H_holo_matr} can be modified to
\begin{equation}
	\addtocounter{equation}{1}
	\mathbf{H}_{\mathrm{mdf}} = \sqrt{\chi_{_\mathrm{0,R}} \chi_{_\mathrm{0,S}}} \mathbf{U}_\mathrm{R} \mathbf{H}_\mathrm{a} \mathbf{U}_\mathrm{S}^\mathrm{T},
	\label{H_em_iso}
\end{equation}
where $\chi_{_\mathrm{0,R}}$ and $\chi_{_\mathrm{0,S}}$ are the antenna efficiency of the receive and source array, respectively. If the antennas are assumed to be lossless, that is, $\chi_{_\mathrm{0,R}} = \chi_{_\mathrm{0,S}} = 1$, \eqref{H_em_iso} reduces to \eqref{H_holo_matr}. It corresponds to the original holographic channel model, which we aim to compare with.

Consider a BS array and a UT array with size $L_{x} = L_{y} = 4 \lambda$. Both arrays have identical element spacings in two dimensions that vary from $\lambda$ to $\lambda/8$, with step size $\lambda/8$. To fully exploit the channel performance, we adopt the water-filling power allocation strategy\cite[Ch.~9]{Thomas2006} to calculate the ergodic capacity
\begin{equation}
	C = \mathbb{E} \left\{ \log _2 \left[ \det \left( \mathbf{I} + \rho (\mathbf{HR_{xx}H} ^\mathrm{H}) \right) \right] \right\},
	\label{cap_erg}
\end{equation}
where $\rho$ is the signal-to-noise ratio (SNR) and $\mathbf{R_{xx}}$ is the autocorrelation matrix of the transmitted signal. In this paper, we set $\rho = 0$ dB, and the total power delivered to the array is $P_\mathrm{total} = 10$ W. Given the channel transfer matrix $\mathbf{H}$ and the total power $P_\mathrm{total}$, $\mathbf{R}_\mathbf{xx}$ can be computed with singular value decomposition precoding.

\begin{figure}[!t]
	\centering
	\subfloat[\label{f8_sm_holo}]{%
		\includegraphics[width=0.45\textwidth]{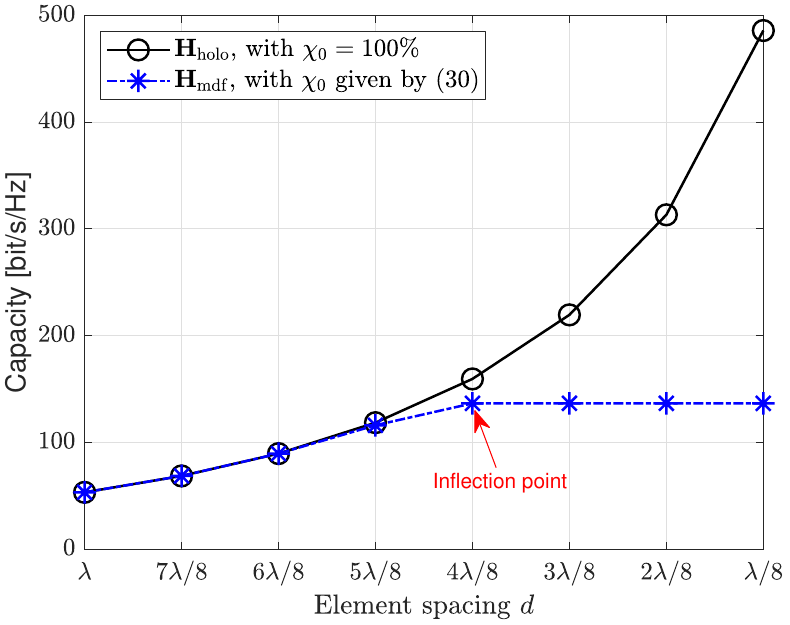}}
	\quad
	\subfloat[\label{f8_cap_xpr}]{%
		\includegraphics[width=0.45\textwidth]{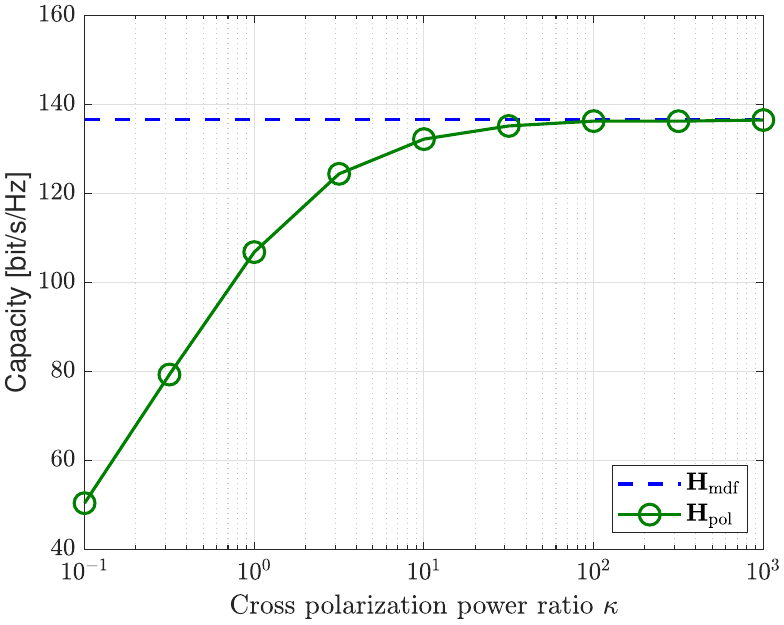}}
		
	\caption{Channel capacity for (a) arrays composed of isotropic antennas under different efficiency considerations, (b) arrays composed of $\boldsymbol{\hat{\theta}}$-polarized isotropic antennas with $\lambda/2$ spacing under different cross-polarization power ratio $\kappa$.}
\end{figure}

To show the results caused by antenna efficiency loss, we adopt channel models \eqref{H_holo_matr} and \eqref{H_em_iso} to calculate the capacity. The results based on a 500-iteration Monte-Carlo simulation are shown in Fig.~\ref{f8_sm_holo}. It is observed that as element spacing gets smaller, the capacity of $\mathbf{H}_\mathrm{holo}$ keeps increasing. However, once the efficiency upper bound is taken into account, the increase in channel performance has an inflection point as the array gets denser. More specifically, the capacity of the modified channel model with efficiency compensation increases with the same rate as the holographic model at the beginning, and then with a slower rate when the element spacing falls in the decreasing region of the element efficiency shown in Fig.~\ref{f7_ele_eff}. Moreover, when the element spacing is smaller than $\lambda/2$, the capacity does not increase. This is because the array gain provided by more elements is offset by the element efficiency loss, and the scaling coefficient due to both factors is kept one, which results in fixed-level eigenvalues.

The performance of the modified model can be predicted using the integral region in Fig.~\ref{f6_crt_pt}. For a square grid array, when the element spacing exceeds $\lambda/\sqrt{2}$, the integral region is totally inside the ellipse (in this case a circle) of the visible region, leading to a 100\% upper bound. As a result, the capacities of both models exhibit the same increasing behavior. However, as the element spacing decreases from $\lambda/\sqrt{2}$ to $\lambda/2$, the ellipse gradually becomes enclosed by the square. Consequently, the performance improvement slows down, and eventually stops, when the visible region is fully within the integral region. This graphical illustration provides a clear understanding of the electromagnetic limitation in holographic communications.

In another case, consider the effect of polarization leakage on channel capacity. Suppose now that the array elements are $\boldsymbol{\hat{\theta}}$-polarized isotropic antennas with $\lambda/2$ spacing. Adopt $\mathbf{H}_\mathrm{pol}$ to calculate the capacity while keeping other settings unchanged. The capacity with different $\kappa$ is given in Fig.~\ref{f8_cap_xpr}. It can be seen that the channel capacity keeps increasing as $\kappa$ becomes larger, and finally tends to be the same as the scalar model case. On the other hand, if $\kappa$ is small, which means the polarization leakage in the environment is severe, the channel capacity decreases rapidly. This simulation result shows how the polarization leakage in the propagation environment influences the performance of holographic MIMO systems.

\subsection{Array of Half-wavelength Dipoles}

\begin{figure}[!t]
	\centering
	\includegraphics[width=0.45\textwidth]{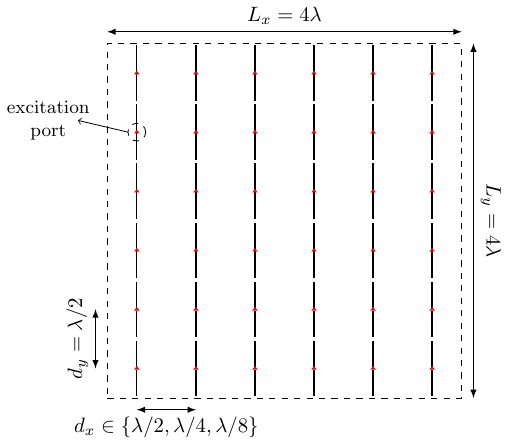}
	\caption{The antenna array composed of half-wavelength dipoles with different element spacings.}
	\label{f9_arr_dip}
\end{figure}

To evaluate the characteristics of the polarized electromagnetic channel model, we use the half-wavelength dipoles as the transceiver elements, which bring us the possibility to shorten the element spacing in one dimension. Fig.~\ref{f9_arr_dip} illustrates the simulation array structure with size $L_x = L_y = 4\lambda$. Firstly, we keep $d_y = \lambda/2$ and change $d_x$ to three different values $\{\lambda/2,\lambda/4,\lambda/8\}$, labeled as $8 \times 8$, $8 \times 16$ and $8 \times 32$, respectively. For each case, we obtain the embedded element patterns for each antenna through electromagnetic simulation, including its $\hat{\boldsymbol{\theta}}$ component and $\hat{\boldsymbol{\phi}}$ component. Finally, we perform a Monte-Carlo simulation and adopt the water-filling algorithm to calculate the ergodic capacity.

For reference, an isolated half-wavelength dipole with resonant frequency $f_\mathrm{r} = 2$ GHz and wavelength $\lambda = 150$ mm is simulated. Table~\ref{tab_dipole} shows its simulation parameters, where $l$ and $r$ are the total length and radius with respect to wavelength $\lambda$. The input impedance $Z_\mathrm{in}$ in this case is a pure real number, which equals to the internal impedance of the voltage source to realize perfect matching.

\begin{figure}[!t]
	\centering
	\begin{minipage}{0.21\textwidth}
		\includegraphics[width=\textwidth]{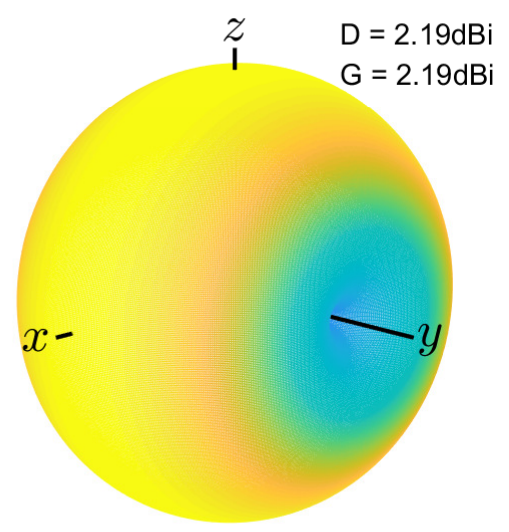}
	\end{minipage}
	\,
	\begin{minipage}{0.21\textwidth}
		\includegraphics[width=\textwidth]{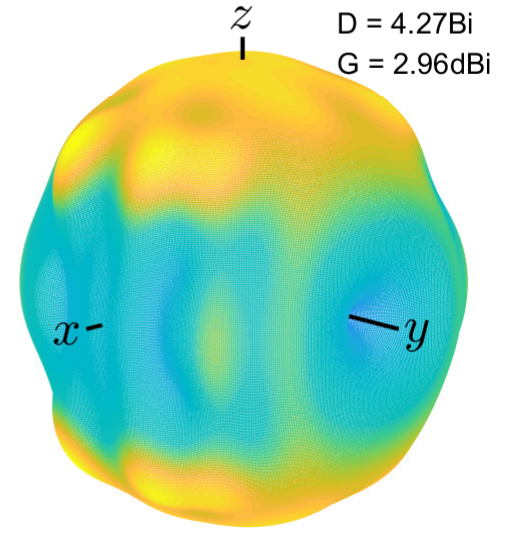}
	\end{minipage}
	\,
	\begin{minipage}{0.21\textwidth}
		\includegraphics[width=\textwidth]{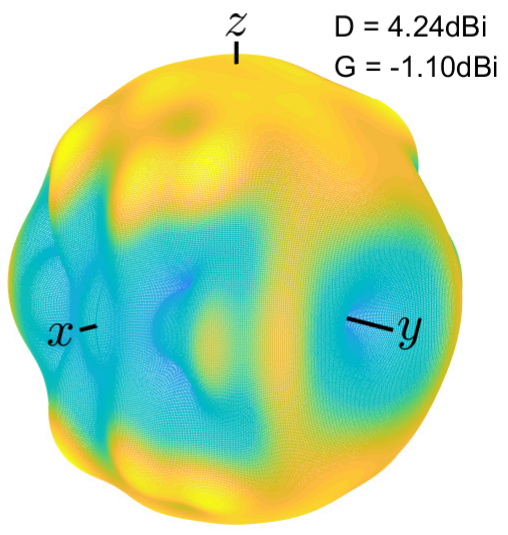}
	\end{minipage}
	\,
	\begin{minipage}{0.21\textwidth}
		\includegraphics[width=\textwidth]{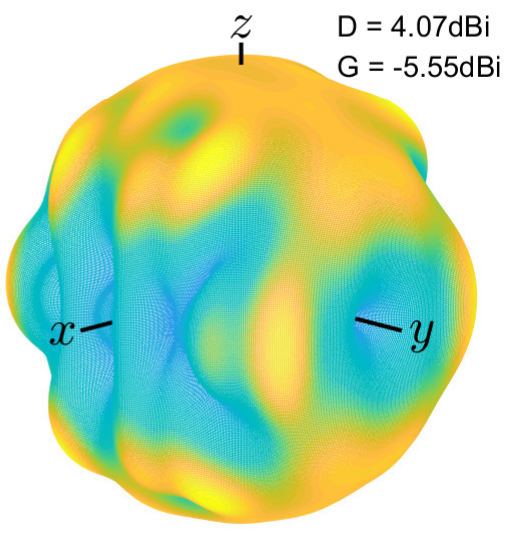}
	\end{minipage}
	\caption{Power patterns of a half-wavelength dipole when it is isolated (the first from the left) and placed at the center of an $8\times8$ array (the second), an $8\times16$ array (the third), and an $8\times32$ array (the fourth).}
	\label{f10_dip_patt}
\end{figure}

Fig.~\ref{f10_dip_patt} shows the power patterns of a half-wavelength dipole when it is isolated and placed in the arrays. It can be observed that mutual coupling leads to the distortion in antenna patterns. From numerical results, the embedded element pattern changes with the location of the antenna. However, the patterns of the central elements are almost the same owing to the similar scattering environment, and their element efficiencies are lower than the theoretical upper bound. Moreover, the elements near the edge suffer from less mutual coupling, leading to higher element efficiencies compared with the upper bound. Table~\ref{tab_array} shows the electromagnetic simulation results of the central antenna elements for different array structures. Here, $\chi_\mathrm{s}$ is the antenna efficiency obtained from electromagnetic simulation, while $\chi_\mathrm{ub}$ is the theoretical upper bound of efficiency in \eqref{eff_e}.

\begin{figure}[!t]
	\centering
	\includegraphics[width=0.45\textwidth]{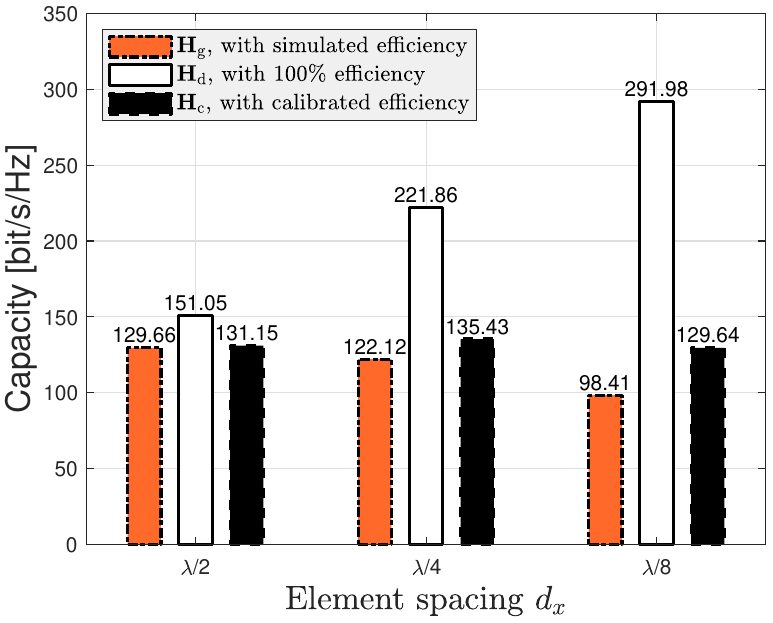}
	\caption{Channel capacity for arrays composed of half-wavelength dipoles under the polarized electromagnetic channel model with different efficiency considerations. The urban street canyon scenario is adopted with $\mu_\mathrm{XPR} = 8.0$ and $\sigma_\mathrm{XPR} = 3.0$.}
	\label{f11_sm_su}
\end{figure}

\begin{table}[!t]
	\renewcommand{\arraystretch}{1.2}
	\centering
	\caption{Simulation parameters of the isolated half-wavelength dipole.}
	\label{tab_dipole}
	\begin{tabular}{ccccc}
		\toprule[1.2pt]
		$f_\mathrm{r}$[GHz] & $\lambda$[m] & $l[\lambda]$ & $r[\lambda]$ & $Z_\mathrm{in}[\Omega]$ \\
		\midrule[0.8pt]
		$2.0$ & $0.15$ & $0.465$ & $0.005$ & $78.3$ \\
		\bottomrule[1.2pt]
	\end{tabular}
\end{table}

\begin{table}[!t]
	\renewcommand{\arraystretch}{1.2}
	\centering
	\caption{Directivity, maximum power gain, simulation efficiency, \\and its theoretical upper bound for the central elements.}
	\label{tab_array}
	\begin{tabular}{ccccc}
		\toprule[1.2pt]
		Structure & $D_{\mathrm{max}}$[dBi] & $G_{\mathrm{max}}$[dBi] & $\chi_\mathrm{s}$[$\%$] & $\chi_\mathrm{ub}$[$\%$] \\
		\midrule[0.8pt]
		dipole (ref) & $2.19$ & $2.19$ & $100$ & $100$ \\
		$8 \times 8$ & $4.27$ & $2.96$ & $74.07$ & $78.54$ \\
		$8 \times 16$ & $4.24$ & $-1.10$ & $29.25$ & $39.27$ \\
		$8 \times 32$ & $4.07$ & $-5.55$ & $10.90$ & $19.63$ \\
		\bottomrule[1.2pt]
	\end{tabular}
\end{table}

In the simulation, each array element is connected to a voltage source with an internal impedance of $78.3~\Omega$. However, because the active impedance of an array element varies due to mutual coupling, there exits impedance mismatch in this setting. Therefore, designing a proper impedance matching network can improve the antenna efficiency. Theoretically, it has been proved that the transmission efficiency upper bound of a phased-array antenna can be reached \cite{Hannan1967}. Moreover, because the influence of mutual coupling mainly comes from the adjacent elements \cite{Craeye2011}, the radiation characteristics of the central elements inside a large-scale array are similar to the infinite case. Nevertheless, the marginal elements in an array suffer from less coupling, and thus their efficiency is larger than the bound. Based on these considerations, we make calibrations to the simulated data as follows:
\begin{equation}
	\chi {_{0,n}} = \max \left(\chi _{\mathrm{s},n},\chi _{\mathrm{ub} ,n}\right), \quad n=1,2,\cdots ,N.
\end{equation}
After the calibration, we can obtain the efficiency matrices $\mathbf{\Gamma}_\mathrm{R}$ and $\mathbf{\Gamma}_\mathrm{S}$ in \eqref{H_em_plr} to get the polarized electromagnetic channel matrix with efficiency calibration, labeled as $\mathbf{H}_{\mathrm{c}}$. To have a better comparison, we consider two other cases to constitute the channel matrix: the first one, denoted by $\mathbf{H}_{\mathrm{g}}$, uses the original simulated data without calibration, the second one, denoted by $\mathbf{H}_{\mathrm{d}}$, considers $100\%$ antenna efficiency for all the elements.

To compare with the simulation results of the isotropic antennas, we perform a Monte-Carlo simulation using \eqref{cap_erg} with $\mathbf{R}_\mathbf{xx}$ given by the water-filling power allocation strategy \cite[Ch.~9]{Thomas2006}. The urban street canyon scenario with $\mu_\mathrm{XPR} = 8.0$ and $\sigma_\mathrm{XPR} = 3.0$ \cite{3GPP38901} is considered. The capacity for different channel models with different element spacings is shown in Fig.~\ref{f11_sm_su}. It can be seen that if we use the original simulated efficiency to constitute the channel matrix $\mathbf{H}_{\mathrm{g}}$, the capacity decreases as the element spacing $d_x$ gets closer. In another case, if we obtain a 100\% antenna efficiency, the capacity calculated with $\mathbf{H}_{\mathrm{d}}$ keeps increasing as the array gets denser. It should be pointed out that the pattern in this case is different from the isolated element, although its efficiency is 100\%. However, the antenna efficiency of the central elements inside an array is limited theoretically by its upper bound. With the calibrated channel model $\mathbf{H}_{\mathrm{c}}$, we can observe that a closer element spacing cannot lead to a boundless capacity improvement. In addition, the cost of the matching network increases as the array gets denser. From an engineering point of view, element spacing below $\lambda /2$ does not contribute more to the communication system.

\subsection{Multi-user Scenario}

\begin{figure*}[!t]
	\centering
	\subfloat[\label{f12_mu_1}]{%
		\includegraphics[width=0.44\linewidth]{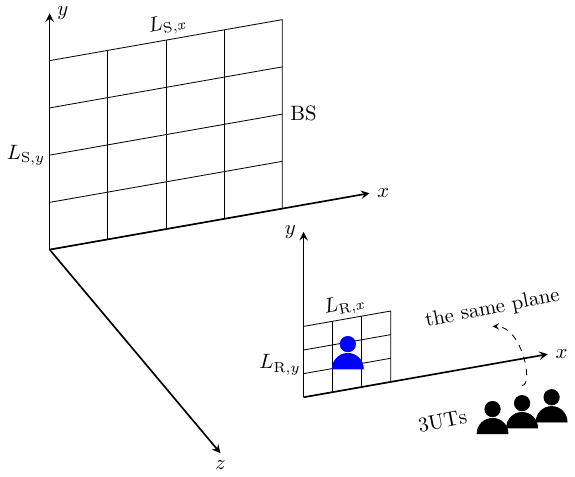}}
	\quad \,
	\subfloat[\label{f12_mu_2}]{%
		\includegraphics[width=0.46\linewidth]{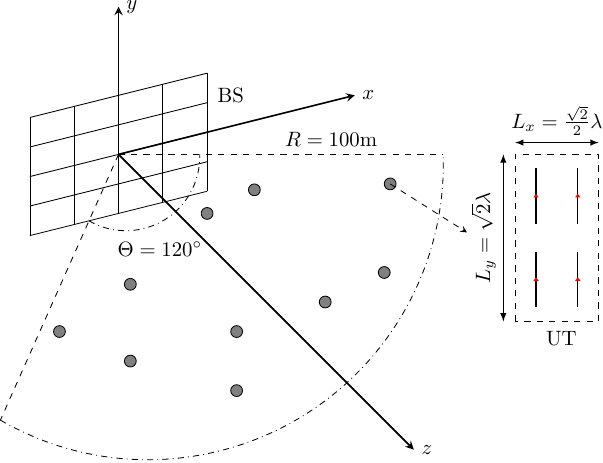}}
	\caption{Multi-user scenarios. (a) Scenario 1, the model in \cite{Wei2022}. (b) Scenario 2, the proposed model.}
	\label{f12_mu_scen}
\end{figure*}

\begin{figure}[!t]
	\centering
	\includegraphics[width=0.45\textwidth]{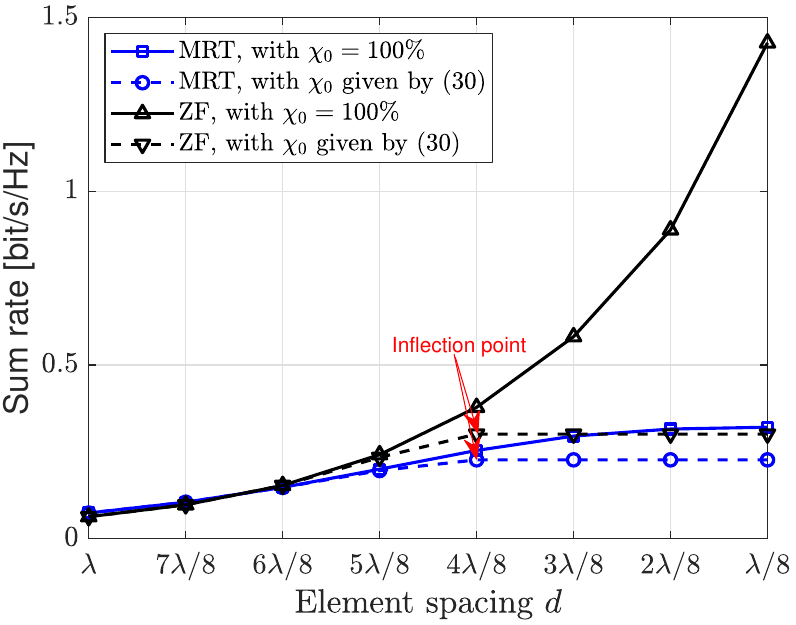}
	\caption{Sum rate in scenario 1, with ZF and MRT pre-coding.}
	\label{f13_sm_mu_1}
\end{figure}

To fully evaluate the holographic system, we perform multi-user simulations based on two different scenarios. In the first case, we adopt the multi-user channel model proposed in \cite{Wei2022} with an identical simulation scenario, as shown in Fig.~\ref{f12_mu_1}. In this model, large-scale fading and mutual coupling are not considered. Arrays in both the BS side and the UT side are composed of isotropic antennas, corresponding to the original holographic channel model \eqref{H_holo_matr}. Three UTs with an array size of $L_{\mathrm{R},x} = L_{\mathrm{R},y} = 4\lambda$ are considered, while the size of the BS array is $L_{\mathrm{S},x} = L_{\mathrm{S},y} = 8\lambda$. All arrays have identical element spacing $d$ that varies from $\lambda$ to $\lambda/8$, with step size $\lambda/8$. SNR is set to 0 dB. To evaluate the effect of mutual coupling, we use \eqref{H_em_iso} to compensate for the antenna efficiency loss. According to the derivation in \cite[Eqs.~(29),(36)]{Wei2022}, the sum rates based on zero-forcing (ZF) and maximum ratio transmission (MRT) precoding schemes are given in Fig.~\ref{f13_sm_mu_1}. It is shown that as the array gets denser, the sum rates of all four cases increase at the beginning, with an identical rate for the respective precoding schemes. However, similar to the single-user case, there is an inflection point at $d=\lambda/2$. The efficiency loss leads to the decrease of SNR, which causes the performance degradation of the system. This conclusion still holds if the large-scale fading is considered, as we will show in the next multi-user scenario.

\begin{figure}[!t]
	\centering
	\includegraphics[width=0.45\textwidth]{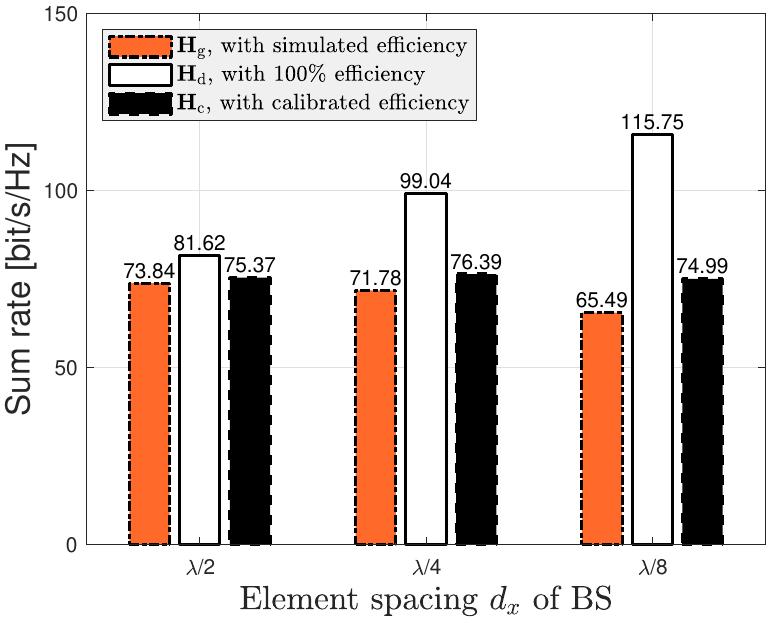}
	\caption{Sum rate in scenario 2, with iterative water-filling.}
	\label{f14_sm_mu_2}
\end{figure}

The second scenario is closer to the real-world case, where the large-scale fading and UT arrays similar to mobile phones are considered. The polarized electromagnetic channel model \eqref{H_em_plr} is adopted to evaluate the system performance. Consider the downlink communication between a dense array BS and $K = 10$ UTs, as shown in Fig.~\ref{f12_mu_2}. The BS and the UTs are arrays composed of half-wavelength dipoles and are all placed in $xy$ plane. The size of the BS array is $4\lambda \times 4\lambda$ with different antenna spacings shown in Fig.~\ref{f9_arr_dip}, while the size of each UT is $\lambda/\sqrt{2} \times \sqrt{2}\lambda$ with a fixed $2 \times 2$ structure shown in Fig.~\ref{f12_mu_2}. UTs are randomly and uniformly distributed in the sector area in front of the BS with $\Theta \in [-60^\circ,60^\circ]$ and $r \in [25,100]$ m. Consider the large-scale fading, the SNR $\rho$ is normalized to $0$ dB at $r = 50$ m. The total power of the BS is $P_\mathrm{total} = 10$ W. The sum rate of this multi-user system is given by the iterative water-filling algorithm \cite{Yu2006}. Fig.~\ref{f14_sm_mu_2} shows the results of the Monte-Carlo simulation. Similar to the single-user scenario, the sum rate calculated with $\mathbf{H}_{\mathrm{d}}$ increases as the BS array gets denser, while that of $\mathbf{H}_{\mathrm{g}}$ decreases. With the calibrated channel model $\mathbf{H}_{\mathrm{c}}$, the sum rates are almost the same in all three cases. This result verifies that antenna efficiency is a key factor which influences the performance of holographic communication. However, compared with the single user scenario, the difference between $\mathbf{H}_{\mathrm{d}}$ and $\mathbf{H}_{\mathrm{c}}$ decreases, this is due to the fact that in scenario 2, each UT is now a $2 \times 2$ array, which weakens the array gain, and the locations of UTs give BS the freedom to allocate the power to achieve the highest sum rate.

\section{Channel Measurement Experiment}

In this section, we perform an indoor channel measurement experiment to derive the channel matrix $\mathbf{H}$ under NLOS scenario. Furthermore, we use the collected data to evaluate the potential of a holographic communication system including possible defects.

\subsection{Experiment Setup and Measurement Results}

\begin{figure}[!t]
	\centering
	\includegraphics[width=0.45\textwidth]{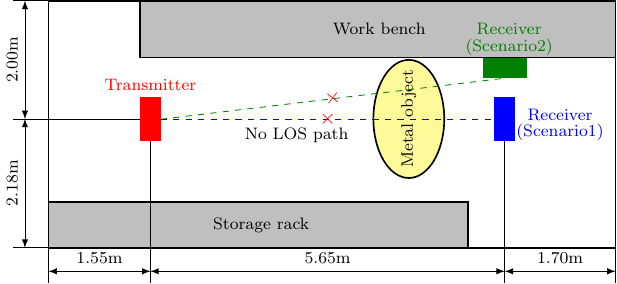}
	\caption{Channel measurement experiment schematic, the LOS path is blocked by the metal object in both scenarios.}
	\label{f15_measure_scen}
\end{figure}

The experiment is performed in a confined space where the LOS path is blocked by a metal object. Many scatterers are present to create a rich scattering environment.
The center frequency $f_\mathrm{c}$ is 4.7 GHz, and the frequency band is 200 MHz with 1023 samples.
The schematic of the measurement environment is shown in Fig.~\ref{f15_measure_scen}.
We conducted two sets of experiments in total. In the first scenario, the virtual receive array plane (explained in the next paragraph) is perpendicular to the transmitter, while in the second scenario, the virtual receive array plane is parallel to the transmitter.

\begin{figure}[!t]
	\centering
	\includegraphics[width=0.45\textwidth]{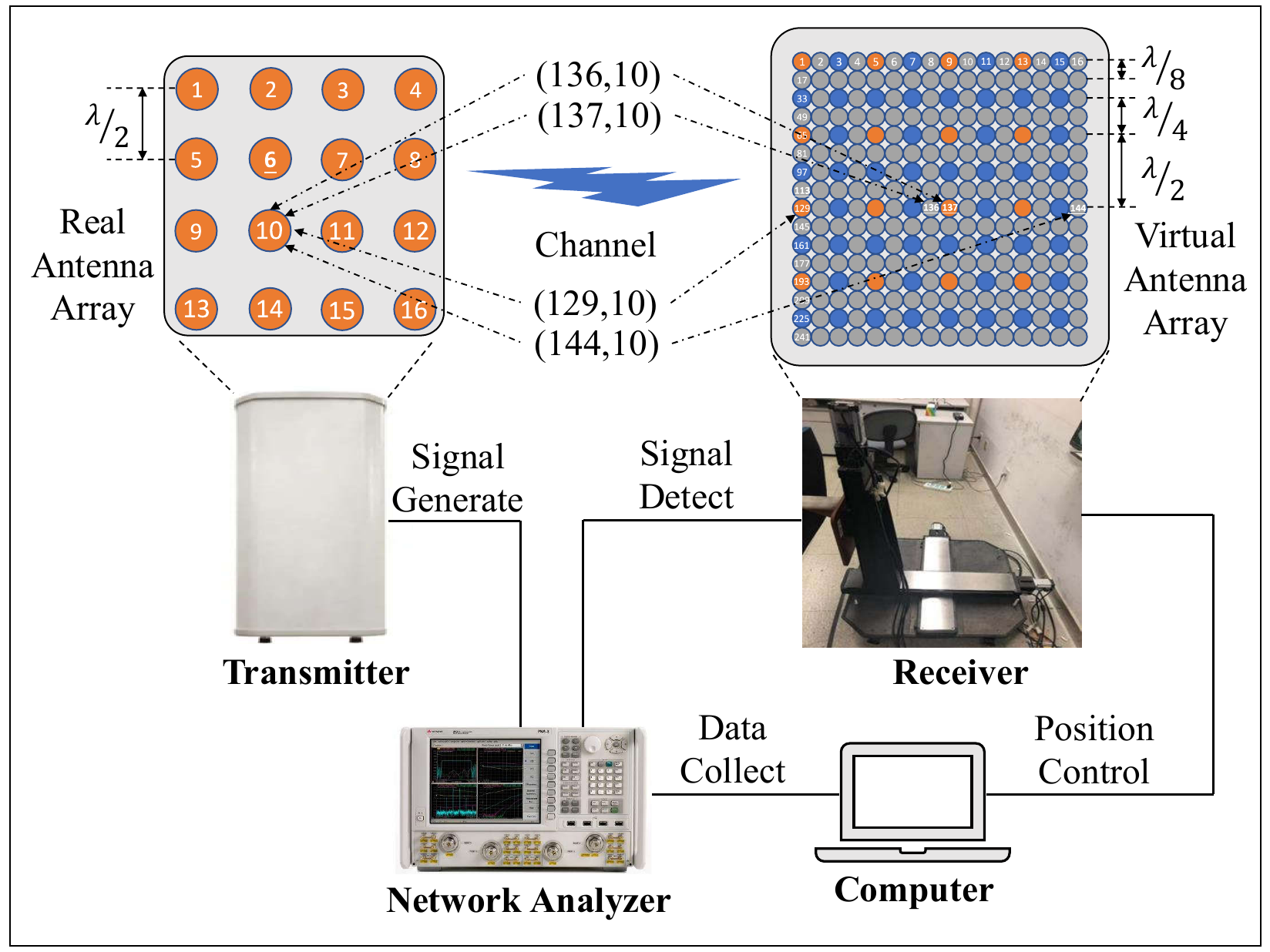}
	\caption{Structures of the channel measurement equipment.}
	\label{f16_measure_diag}
\end{figure}

As shown in Fig.~\ref{f16_measure_diag}, the source array is a real array containing $N_{\mathrm{S}} = 16$ antenna elements, and is composed of patch antennas whose half power beam width is 70$^{\circ}$.
On the receiver side, a single discone antenna is used to achieve an omnidirectional pattern.
The receive antenna is controlled by electric machines and can be moved onto different positions in a plane to construct a virtual dense array with arbitrary element spacing.
In each round, we excite a different element of the source array, move the receiver to $N_{\mathrm{R}} = 256$ preset positions.
The distance of two adjacent preset positions is $\lambda /8$.
Upon each measurement, we use a calibrated network analyzer to obtain the S parameter between a source antenna and a receive antenna, which can be regarded as one entry of the channel matrix $\mathbf{H}$ \cite{Michel2010}.
In this way, the channel matrix $\mathbf{H}$ can be obtained element by element. During the measurement, the environment of the indoor laboratory keeps unchanged, and therefore we assume the propagation environment is stationary.

Each transceiver element pair $(q,p)$ represents the relationship between the $q$-th antenna element in the virtual receive array and the $p$-th antenna element in the transmit array.
After measurement, a channel matrix $\mathbf{H}$ with size $N_{\mathrm{R}} \times N_{\mathrm{S}}$ is obtained.
If we extract the rows of $\mathbf{H}$ regularly, e.g., using the rows corresponding to the antenna elements with $\lambda /2$ spacing instead of $\lambda /8$, a new channel matrix with a larger element spacing can be obtained.

\begin{figure}[!t]
	\centering
	\subfloat[\label{f17_S_a}]{%
		\includegraphics[width=0.43\textwidth]{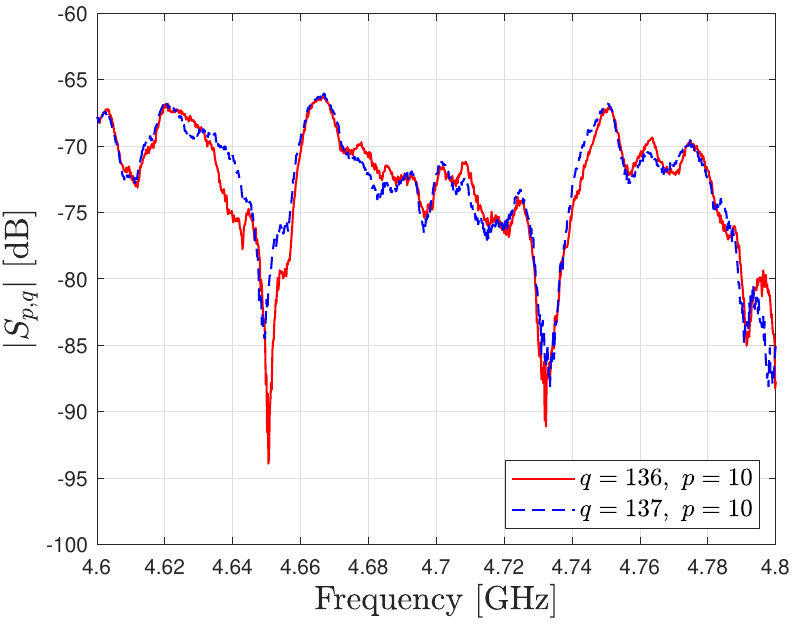}}
	\quad
	\subfloat[\label{f17_S_b}]{%
		\includegraphics[width=0.43\textwidth]{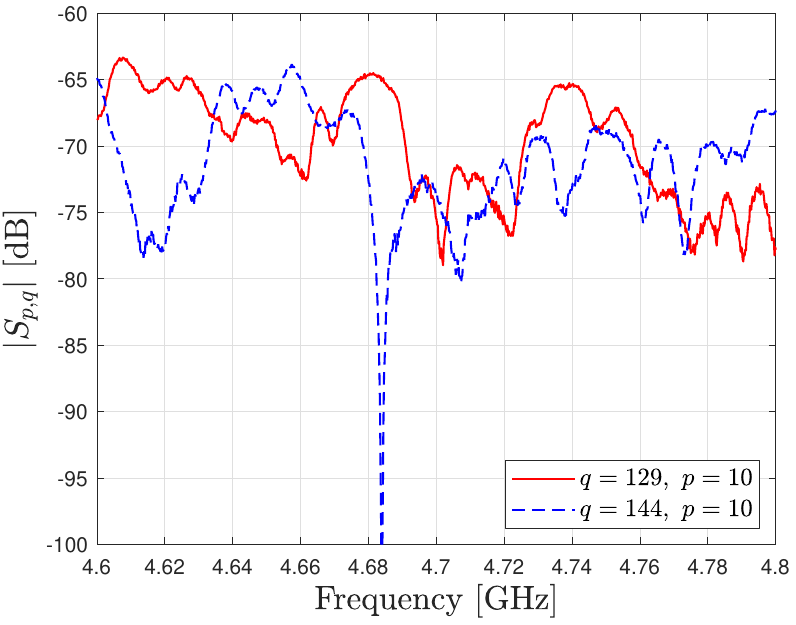}}
	\caption{Channel responses at different receive elements when the transmit element is fixed. The two receive elements are (a) separated by $\lambda /8$. (b) separated by $2 \lambda$.}
	\label{f17_S_param} 
\end{figure}

The responses corresponding to (136,10) and (137,10) transceiver pairs in the second scenario are shown in Fig.~\ref{f17_S_a}.
Since the receive antennas are adjacent, we can observe that the channel responses are quite similar, which verifies that the propagation environment is stationary.
On the other hand, the channel responses for the transceiver pairs (129,10) and (144,10) are shown in Fig.~\ref{f17_S_b}. Because the antenna elements are separated by $2 \lambda$, their channel responses are different in the whole spectrum, showing a low correlation compared with the results in Fig.~\ref{f17_S_a}. The difference between these two figures shows the effect of spatial coherence.

\subsection{Channel Performance Evaluation}

\begin{figure}[!t]
	\centering
	\subfloat[\label{f18_C_a}]{%
		\includegraphics[width=0.45\textwidth]{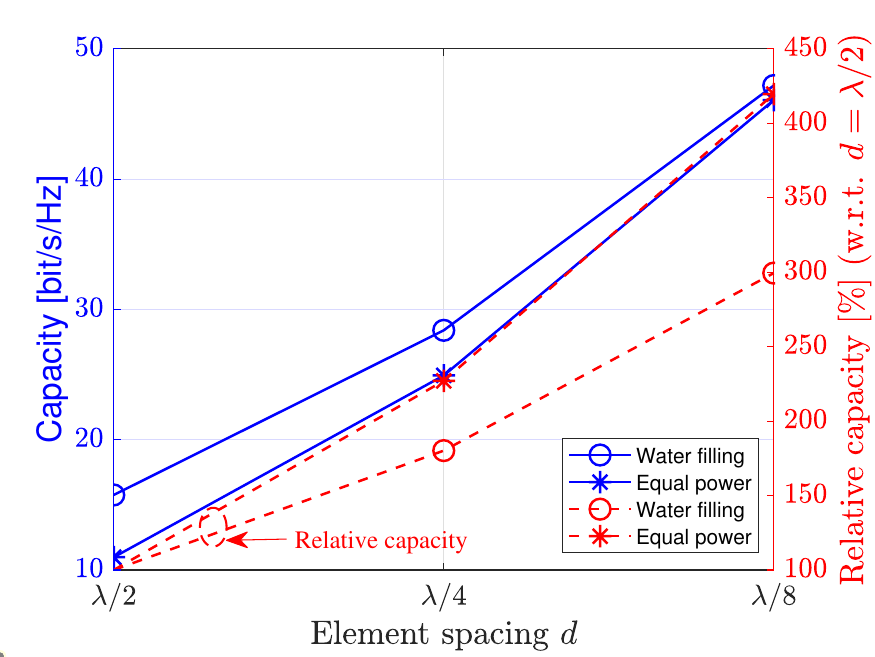}}
	\quad
	\subfloat[\label{f18_C_b}]{%
		\includegraphics[width=0.45\textwidth]{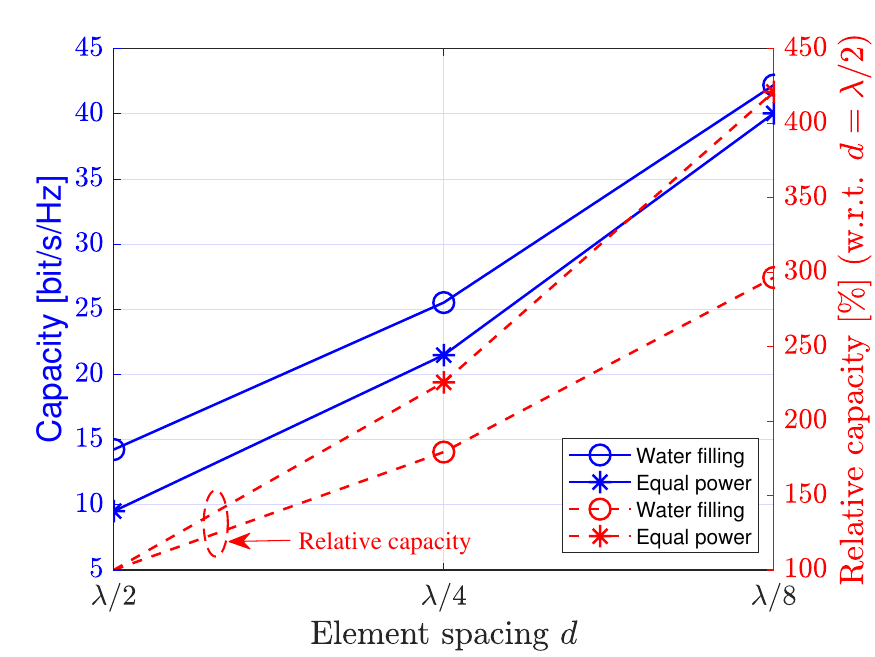}}
	\caption{Channel capacity and relative capacity with different virtual receive array. (a) Scenario 1. (b) Scenario 2.}
	\label{f18_cap_ideal} 
\end{figure}

Once the channel matrix $\mathbf{H}$ is obtained, we can evaluate the channel performance. Firstly, we resample two sub-matrices, corresponding to a receive array with element spacings $\lambda/4$ and $\lambda/2$, respectively. The total antenna number of the virtual receive array with element spacing $d \in \{\lambda/8, \lambda/4, \lambda/2\}$ are $N_\mathrm{R} \in \{256, 64, 16\}$. The variation of capacity with different element spacings in both scenarios is shown in Fig.~\ref{f18_cap_ideal}. Both the water-filling power allocation and the equal power allocation strategies are used to evaluate the channel performance. Solid lines correspond to the channel capacity while dashed lines correspond to the relative capacity with respect to the capacity obtained with $\lambda /2$ spacing in the receiver side.

\begin{figure}[!t]
	\centering
	\subfloat[\label{f16_C_c}]{%
		\includegraphics[width=0.46\textwidth]{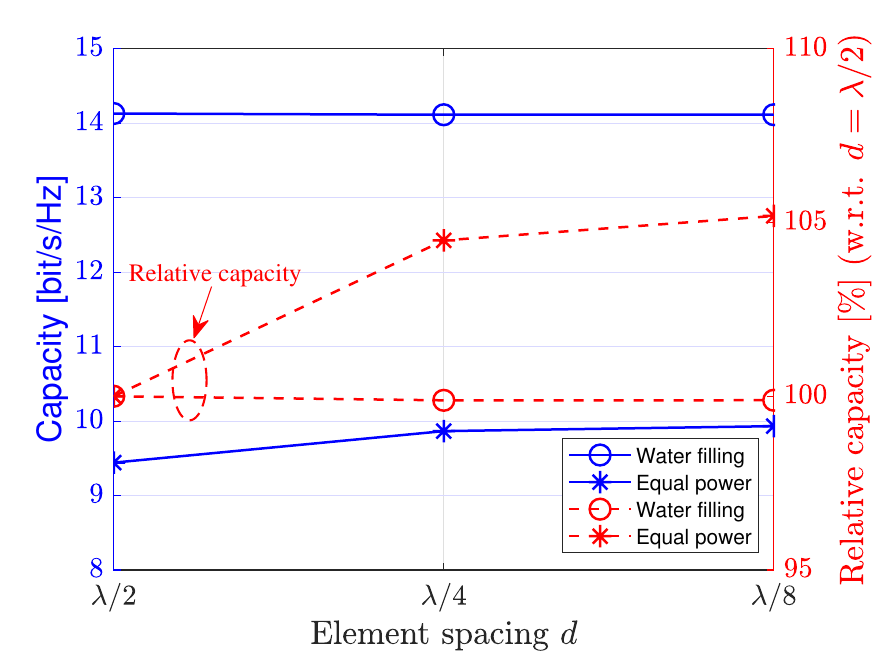}}
	\quad
	\subfloat[\label{f16_C_d}]{%
		\includegraphics[width=0.46\textwidth]{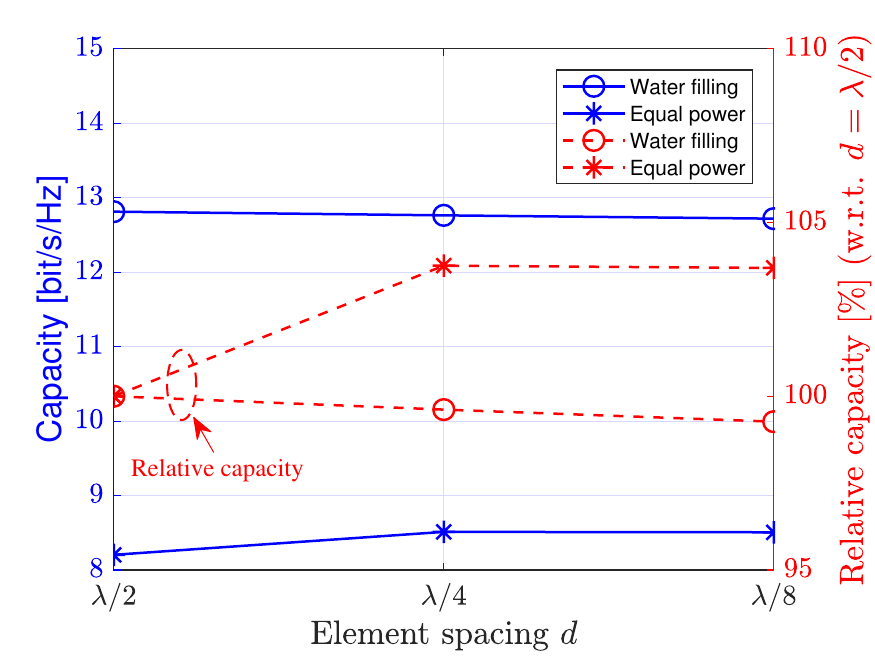}}
	\caption{Channel capacity and relative capacity with efficiency compensation. (a) Scenario 1. (b) Scenario 2.}
	\label{f19_cap_real} 
\end{figure}

It can be seen that given a limited space, a more compact structure on the receiver side appears to yield higher capacity. The capacity seems to increase boundlessly as the array structure gets denser, similar to the results obtained from the holographic channel model. This result, however, is rational only if the coupling in the receiver side is ignored, as this virtual receive array is constructed by a single antenna, maintaining its isolated pattern and input impedance. Consequently, we should consider this non-ideal factor in the data post-processing stage.

The effect of mutual coupling can be divided into two terms, the first is the distortion of the embedded element pattern, and the second is the degradation of antenna efficiency. In large-scale arrays, the patterns of the elements inside the array are almost the same\cite[Ch.~4]{Arun2005}, because dummy elements with no excitation are usually placed near the marginal elements to provide similar scattering environment \cite{Zhu2022}. Therefore, we make a compensation only for the antenna efficiency. The modified channel matrix is given by $\mathbf{H}^\prime = \sqrt{\chi_{_\mathrm{0,R}} \chi_{_\mathrm{0,S}}} \mathbf{H}$, where $\chi_{_\mathrm{0,R}}$ and $\chi_{_\mathrm{0,S}}$ are obtained from theoretical calculation. The simulation results based on the calibrated data are shown in Fig.~\ref{f19_cap_real}. Compared with Fig.~\ref{f18_cap_ideal}, the variation of channel capacity is very small for different receive arrays. This implies that once the electromagnetic limitation is taken into account, a more compact structure in a given space does not lead to an unbounded increase of channel performance. Typically, an element spacing of $\lambda /2$ is a good choice to design a square grid antenna array.

\begin{figure}[!t]\centering
	\centering
	\subfloat[\label{f20_eig_a}]{%
		\includegraphics[width=0.43\textwidth]{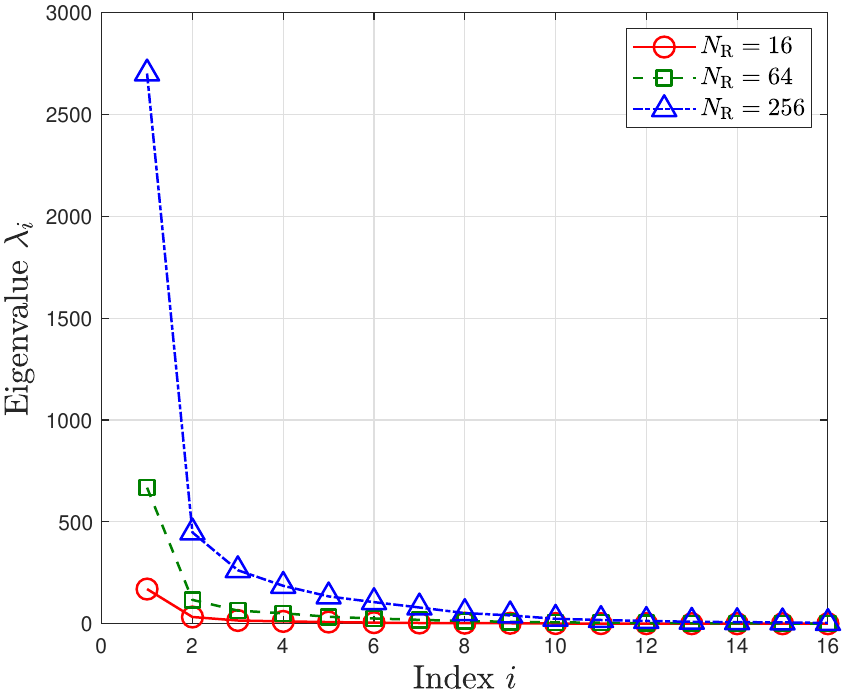}}
	\quad
	\subfloat[\label{f20_eig_b}]{%
		\includegraphics[width=0.43\textwidth]{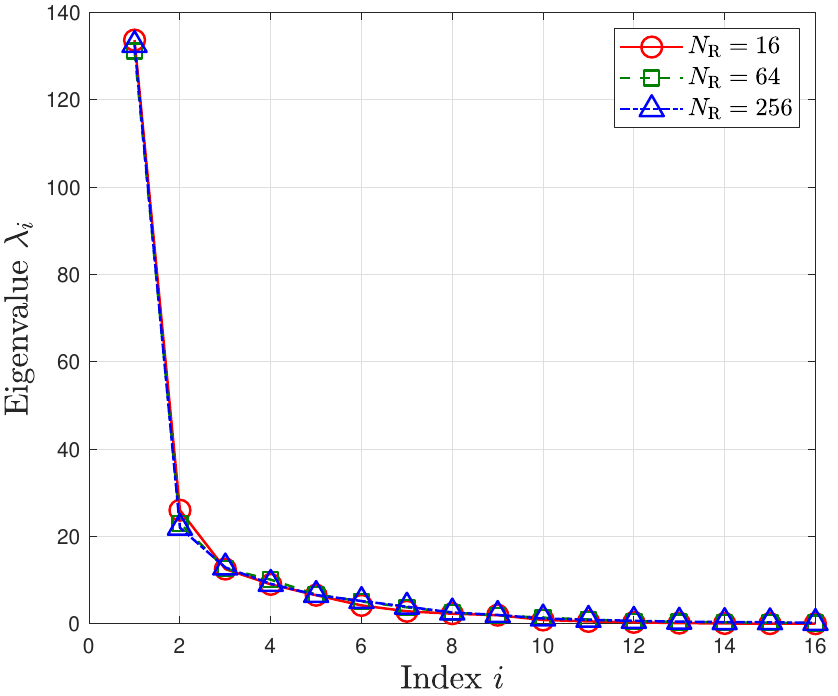}}
	\caption{Eigenvalues of $\mathbf{HH^\mathrm{H}}$ in scenario 2. (a) From direct measurement. (b) With efficiency compensation.}
	\label{f20_ex_eig}
\end{figure}

The variation of capacity can be understood from the perspective of eigenvalues. Fig.~\ref{f20_ex_eig} shows the first 16 eigenvalues of $\mathbf{HH}^\mathrm{H}$ with and without calibration in scenario 2. It is shown that without mutual coupling, the eigenvalues corresponding to communication sub-channels get larger as the element number increases. However, with efficiency compensation, the eigenvalues of different receive array structures tend to be the same. This compensation, however, gives an approximation since we ignore the distortion of antenna patterns due to mutual coupling, which usually leads to performance degradation.

\section{Conclusion and Discussion}
In this paper, we proposed an electromagnetic channel model for holographic MIMO systems. The new model incorporates the radiation and polarization characteristics of electromagnetic waves. The study of coupling and radiation theories reveals that there exist performance constraints in holographic communications. With numerical simulation results and a channel measurement experiment, it is demonstrated that antenna efficiency is the most crucial challenge for capacity improvement. Therefore, additional attention should be paid to the electromagnetic constraints within dense antenna arrays in holographic communications.

It is worthwhile to point out that in a finite antenna array, elements located near the edge suffer from less coupling. Furthermore, the dense array antennas can also have an irregular lattice, which may improve the performance limit of the central elements. Further research could focus on these fields to provide overall guidance for holographic communications.

\section*{Acknowledgment}
\addcontentsline{toc}{section}{Acknowledgment}
The authors would like to thank the anonymous reviewers and editors for their time and effort, which greatly improves the quality of this paper.

\bibliographystyle{IEEEtran}
\bibliography{bib_book,bib_paper}

\begin{thebibliography}{10}
\providecommand{\url}[1]{#1}
\csname url@samestyle\endcsname
\providecommand{\newblock}{\relax}
\providecommand{\bibinfo}[2]{#2}
\providecommand{\BIBentrySTDinterwordspacing}{\spaceskip=0pt\relax}
\providecommand{\BIBentryALTinterwordstretchfactor}{4}
\providecommand{\BIBentryALTinterwordspacing}{\spaceskip=\fontdimen2\font plus
\BIBentryALTinterwordstretchfactor\fontdimen3\font minus
  \fontdimen4\font\relax}
\providecommand{\BIBforeignlanguage}[2]{{%
\expandafter\ifx\csname l@#1\endcsname\relax
\typeout{** WARNING: IEEEtran.bst: No hyphenation pattern has been}%
\typeout{** loaded for the language `#1'. Using the pattern for}%
\typeout{** the default language instead.}%
\else
\language=\csname l@#1\endcsname
\fi
#2}}
\providecommand{\BIBdecl}{\relax}
\BIBdecl

\bibitem{Chettri2020}
L.~Chettri and R.~Bera, ``A comprehensive survey on internet of things {(IoT)}
  toward {5G} wireless systems,'' \emph{{IEEE }Internet Things J.}, vol.~7,
  no.~1, pp. 16--32, Jan. 2020.

\bibitem{Boccardi2014}
F.~Boccardi, R.~W. Heath, A.~Lozano, T.~L. Marzetta, and P.~Popovski, ``Five
  disruptive technology directions for 5{G},'' \emph{{IEEE} Commun. Mag.},
  vol.~52, no.~2, pp. 74--80, Feb. 2014.

\bibitem{3GPP38901}
3GPP, ``{TR} 38.901 v16.1.0, {Study} on channel model for frequencies from 0.5
  to 100 {GHz},'' {3rd Generation Partnership Project (3GPP)}, Tech. Rep., Dec.
  2019.

\bibitem{Marzetta2010}
T.~L. Marzetta, ``Noncooperative cellular wireless with unlimited numbers of
  base station antennas,'' \emph{{IEEE} Trans. Wirel. Commun.}, vol.~9, no.~11,
  pp. 3590--3600, Nov. 2010.

\bibitem{Larsson2014}
E.~G. Larsson, O.~Edfors, F.~Tufvesson, and T.~L. Marzetta, ``Massive {MIMO}
  for next generation wireless systems,'' \emph{{IEEE} Commun. Mag.}, vol.~52,
  no.~2, pp. 186--195, Feb. 2014.

\bibitem{Swindlehurst2014}
A.~L. Swindlehurst, E.~Ayanoglu, P.~Heydari, and F.~Capolino, ``Millimeter-wave
  massive {MIMO}: the next wireless revolution?'' \emph{{IEEE} Commun. Mag.},
  vol.~52, no.~9, pp. 56--62, Sep. 2014.

\bibitem{Lu2014}
L.~Lu, G.~Y. Li, A.~L. Swindlehurst, A.~Ashikhmin, and R.~Zhang, ``An overview
  of massive {MIMO}: Benefits and challenges,'' \emph{{IEEE} J. Sel. Top.
  Signal Process.}, vol.~8, no.~5, pp. 742--758, Oct. 2014.

\bibitem{Bjornson2017}
E.~Bj\"{o}rnson, J.~Hoydis, and L.~Sanguinetti, ``Massive {MIMO} networks:
  Spectral, energy, and hardware efficiency,'' \emph{Foundations and
  Trends$^\circledR$ in Signal Processing}, vol.~11, no. 3--4, pp. 154--655,
  2017.

\bibitem{Loyka2004}
S.~Loyka, ``Information theory and electromagnetism: Are they related?'' in
  \emph{{Proc. 10th Int. Symp. Antenna Technol. Appl. Electromagn. URSI
  Conf.}}, 2004, pp. 1--5.

\bibitem{Ivrlac2014}
M.~T. Ivrlac and J.~A. Nossek, ``The multiport communication theory,''
  \emph{{IEEE} Circuits Syst. Mag.}, vol.~14, no.~3, pp. 27--44, Aug. 2014.

\bibitem{Franceschetti2018}
M.~Franceschetti, \emph{Wave Theory of Information}.\hskip 1em plus 0.5em minus
  0.4em\relax Cambridge University Press, 2018.

\bibitem{Wu2019}
Q.~Wu and R.~Zhang, ``Intelligent reflecting surface enhanced wireless network
  via joint active and passive beamforming,'' \emph{{IEEE} Trans. Wireless
  Commun.}, vol.~18, no.~11, pp. 5394--5409, Nov. 2019.

\bibitem{Haraz2014}
O.~M. Haraz, A.~Elboushi, S.~A. Alshebeili, and A.-R. Sebak, ``Dense dielectric
  patch array antenna with improved radiation characteristics using {EBG}
  ground structure and dielectric superstrate for future {5G} cellular
  networks,'' \emph{IEEE Access}, vol.~2, pp. 909--913, Aug. 2014.

\bibitem{Pizzo2020Degrees}
A.~Pizzo, T.~L. Marzetta, and L.~Sanguinetti, ``Degrees of freedom of
  holographic {MIMO} channels,'' in \emph{2020 IEEE 21st International Workshop
  on Signal Processing Advances in Wireless Communications (SPAWC)}, 2020, pp.
  1--5.

\bibitem{Pizzo2022Nyquist}
A.~Pizzo, A.~d.~J. Torres, L.~Sanguinetti, and T.~L. Marzetta, ``Nyquist
  sampling and degrees of freedom of electromagnetic fields,'' \emph{{IEEE}
  Trans. Signal Processing}, vol.~70, pp. 3935--3947, Jun. 2022.

\bibitem{Pizzo2020Spatially}
A.~Pizzo, T.~L. Marzetta, and L.~Sanguinetti, ``Spatially-stationary model for
  holographic {MIMO} small-scale fading,'' \emph{{IEEE} J. Select. Areas
  Commun.}, vol.~38, no.~9, pp. 1964--1979, Sep. 2020.

\bibitem{Gong2023}
T.~Gong, I.~Vinieratou, R.~Ji, H.~Chongwen, G.~Alexandropoulos, L.~Wei,
  Z.~Zhang, M.~Debbah, H.~V. Poor, and C.~Yuen, ``Holographic {MIMO}
  communications: Theoretical foundations, enabling technologies, and future
  directions,'' arXiv, 2023. [Online]. Available:
  https://arxiv.org/abs/2212.01257.

\bibitem{Pizzo2022Fourier}
A.~Pizzo, L.~Sanguinetti, and T.~L. Marzetta, ``Fourier plane-wave series
  expansion for holographic {MIMO} communications,'' \emph{{IEEE} Trans.
  Wireless Commun.}, vol.~21, no.~9, pp. 6890--6905, Sep. 2022.

\bibitem{Pizzo2022Spatial}
------, ``Spatial characterization of electromagnetic random channels,''
  \emph{{IEEE} Open J. Commun. Soc.}, vol.~3, pp. 847--866, Apr. 2022.

\bibitem{Sun2022}
\BIBentryALTinterwordspacing
S.~Sun and M.~Tao, ``Characteristics of channel eigenvalues and mutual coupling
  effects for holographic reconfigurable intelligent surfaces,''
  \emph{Sensors}, vol.~22, no.~14, Jul. 2022. [Online]. Available:
  \url{https://www.mdpi.com/1424-8220/22/14/5297}
\BIBentrySTDinterwordspacing

\bibitem{Yuan2023}
S.~S.~A. Yuan, X.~Chen, C.~Huang, and W.~E.~I. Sha, ``Effects of mutual
  coupling on degree of freedom and antenna efficiency in holographic {MIMO}
  communications,'' \emph{{IEEE} Open J. Antennas Propag.}, vol.~4, pp.
  237--244, Feb. 2023.

\bibitem{Wei2022}
L.~Wei, C.~Huang, G.~C. Alexandropoulos, W.~E.~I. Sha, Z.~Zhang, M.~Debbah, and
  C.~Yuen, ``Multi-user holographic {MIMO} surfaces: Channel modeling and
  spectral efficiency analysis,'' \emph{{IEEE} J. Sel. Top. Sign. Proces.},
  vol.~16, no.~5, pp. 1112--1124, Aug. 2022.

\bibitem{Sha2022}
L.~Wei, C.~Huang, G.~C. Alexandropoulos, Z.~Yang, J.~Yang, W.~E.~I. Sha,
  Z.~Zhang, M.~Debbah, and C.~Yuen, ``Tri-polarized holographic {MIMO} surface
  in near-field: Channel modeling and precoding design,'' arXiv, 2023.
  [Online]. Available: https://arxiv.org/abs/2211.03479.

\bibitem{Akrout2023}
M.~Akrout, V.~Shyianov, F.~Bellili, A.~Mezghani, and R.~W. Heath,
  ``Super-wideband massive {MIMO},'' \emph{{IEEE} J. Sel. Areas Commun.},
  vol.~41, no.~8, pp. 2414--2430, Aug. 2023.

\bibitem{Yongxi2022}
Y.~Liu, M.~Zhang, and T.~Wang, ``Effect of antenna pattern on the
  electromagnetic {MIMO} communication,'' in \emph{{IEEE} International
  Conference on Communication Technology (ICCT)}, Nanjing, China, Nov. 2022.

\bibitem{Tengjiao2022}
T.~Wang, W.~Han, Z.~Zhong, J.~Pang, G.~Zhou, S.~Wang, and Q.~Li,
  ``Electromagnetic-compliant channel modeling and performance evaluation for
  holographic {MIMO},'' in \emph{{IEEE} Global Communications Conference
  (GLOBECOM)}, Rio de Janeiro, Brazil, Dec. 2022.

\bibitem{Prabhakar2022}
P.~H. Pathak and R.~J. Burkholder, \emph{Electromagnetic Radiation, Scattering,
  and Diffraction}.\hskip 1em plus 0.5em minus 0.4em\relax John Wiley \& Sons,
  2022.

\bibitem{Pozar1994}
D.~Pozar, ``The active element pattern,'' \emph{{IEEE} Trans. Antennas
  Propag.}, vol.~42, no.~8, pp. 1176--1178, Aug. 1994.

\bibitem{Arun2005}
A.~K. Bhattacharyya, \emph{Phased Array Antennas: Floquet Analysis, Synthesis,
  BFNs and Active Array Systems}, 1st~ed.\hskip 1em plus 0.5em minus
  0.4em\relax John Wiley \& Sons, 2005.

\bibitem{Pozar2012}
D.~M. Pozar, \emph{Microwave Engineering}, 4th~ed.\hskip 1em plus 0.5em minus
  0.4em\relax John Wiley \& Sons, 2012.

\bibitem{Craeye2011}
C.~Craeye and D.~González-Ovejero, ``A review on array mutual coupling
  analysis,'' \emph{Radio Sci.}, vol.~46, no.~02, pp. 1--25, Apr. 2011.

\bibitem{Balanis2016}
C.~A. Balanis, \emph{Antenna Theory: Analysis and Design}, 4th~ed.\hskip 1em
  plus 0.5em minus 0.4em\relax John Wiley \& Sons, 2016.

\bibitem{Vandenbosch2010}
G.~A.~E. Vandenbosch, ``Reactive energies, impedance, and {$Q$} factor of
  radiating structures,'' \emph{{IEEE} Trans. Antennas Propag.}, vol.~58,
  no.~4, pp. 1112--1127, Apr. 2010.

\bibitem{Shahpari2018}
M.~Shahpari and D.~V. Thiel, ``Fundamental limitations for antenna radiation
  efficiency,'' \emph{{IEEE} Trans. Antennas Propag.}, vol.~66, no.~8, pp.
  3894--3901, Aug. 2018.

\bibitem{Hannan1964}
P.~W. Hannan, ``The element-gain paradox for a phased-array antenna,''
  \emph{{IEEE} Trans. Antennas Propag.}, vol.~12, no.~4, pp. 423--433, Jul.
  1964.

\bibitem{Kahn1967}
W.~Kahn, ``Ideal efficiency of a radiating element in an infinite array,''
  \emph{{IEEE} Trans. Antennas Propag.}, vol.~15, no.~4, pp. 534--538, Jul.
  1967.

\bibitem{Kildal2016}
P.~S. Kildal, A.~Vosoogh, and S.~Maci, ``Fundamental directivity limitations of
  dense array antennas: A numerical study using {Hannan}’s embedded element
  efficiency,'' \emph{{IEEE} Antennas Wireless Propag. Lett.}, vol.~15, pp.
  766--769, Aug. 2016.

\bibitem{Oestges2008}
C.~Oestges, B.~Clerckx, M.~Guillaud, and M.~Debbah, ``Dual-polarized wireless
  communications: from propagation models to system performance evaluation,''
  \emph{{IEEE} Trans. Wireless Commun.}, vol.~7, no.~10, pp. 4019--4031, Oct.
  2008.

\bibitem{Thomas2006}
T.~M. Cover and J.~A. Thomas, \emph{Elements of Information Theory}.\hskip 1em
  plus 0.5em minus 0.4em\relax John Wiley \& Sons, 2006.

\bibitem{Hannan1967}
P.~W. Hannan, ``Proof that a phased-array antenna can be impedance matched for
  all scan angles,'' \emph{Radio Sci.}, vol.~2, no.~3, pp. 361--371, Mar. 1967.

\bibitem{Yu2006}
W.~Yu, ``Sum-capacity computation for the gaussian vector broadcast channel via
  dual decomposition,'' \emph{{IEEE} Trans. Inf. Theory}, vol.~52, no.~2, pp.
  754--759, Feb. 2006.

\bibitem{Michel2010}
M.~T. Ivrlač and J.~A. Nossek, ``Toward a circuit theory of communication,''
  \emph{{IEEE} Trans. Circuits Syst. I, Reg. Papers1}, vol.~57, no.~7, pp.
  1663--1683, Jul. 2010.

\bibitem{Zhu2022}
Q.~Zhu, J.~Fang, X.~Zhang, and M.~Jin, ``Influence of dummy elements on the
  performance of array antennas,'' in \emph{{IEEE} 10th Asia-Pacific Conference
  on Antennas and Propagation (APCAP)}, 2022, pp. 1--2.

\end{thebibliography}

\vfill

\end{document}